%% file: main.tex
\let\oldvec\vec
\documentclass[runningheads]{llncs}
\let\vec\oldvec

\input{structure.tex}

\begin{document}

\title{Normalising Lustre Preserves Security}
\institute{Indian Institute of Technology Delhi, INDIA
\email{\{sanjiva,madhukar.yr\}@cse.iitd.ac.in}}
\author{Sanjiva Prasad 
\and 
R. Madhukar Yerraguntla
}


\maketitle

\begin{abstract}
The synchronous reactive data flow language \lustre\ is an expressive language, equipped with a suite of tools for modelling, simulating and model-checking a wide variety of safety-critical systems.
A critical intermediate step in the formally certified compilation of \lustre\ involves translation to a well-behaved sub-language called ``Normalised \lustre'' (\nlustre).
Recently, we proposed a simple Denning-style lattice-based secure information flow type system for \nlustre, and proved its soundness by establishing that security-typed programs are non-interfering with respect to the co-inductive stream semantics.

In this paper, we propose a similar security type system for unrestricted \lustre, and show that Bourke \textit{et al.}'s semantics-preserving normalisation transformations from \lustre\ to \nlustre\ are security-preserving as well.  
A novelty is the use of refinement security types for node calls.
The main result is the preservation of security types by the normalisation transformations. 
The soundness of our security typing rules is shown by establishing that well-security-typed programs are non-interfering, via a reduction to type-preservation (here), semantics-preservation (Bourke \textit{et al.}) and our previous result of non-interference for \nlustre. \\

\lustre,
Compiler transformation,
Security type system,
Non-interference,
Security preservation.
\end{abstract}

\input{intro.tex}

\input{LustreSyntax.tex}

\input{LustreSecTyping.tex}
\input{normalisation.tex}

\input{noninterference.tex}





\input{conclusions}

\bibliographystyle{acm}
\bibliography{bibliography}
\newpage
\appendix
\input{APP_free_vars.tex}
\input{APP_consolidated_stream_semantics.tex}

\input{APP_aux_pred.tex}
\input{running_example.tex}

\end{document}

%% file: structure.tex
\usepackage{graphicx,lipsum}
\usepackage{algorithm,algpseudocode}
\usepackage[hyphens,spaces,obeyspaces]{url}
\usepackage[dvipsnames]{xcolor}
\usepackage{figlatex,wrapfig}
\usepackage{listings,amssymb,mathtools}
\usepackage{stmaryrd}
\usepackage{mathrsfs}
\usepackage{array,multirow}
\usepackage{pifont}
\usepackage{wrapfig}
\usepackage{textcomp}
\usepackage{xspace}
\usepackage{trfrac}
\usepackage{amsmath}
\usepackage{afterpage}
\usepackage[colorlinks]{hyperref}
\usepackage{tcolorbox}
\usepackage[T1]{fontenc}
\usepackage[utf8]{inputenc} 
\usepackage{tikz}
\usepackage{float}
\usepackage{esvect}
\usepackage{verbatim}
\usepackage{stmaryrd}
\usepackage{turnstile}
\usepackage{xcolor}
\usepackage{soul}

\definecolor{eclipseBlue}{RGB}{42,0.0,255}
\definecolor{eclipseGreen}{RGB}{63,127,95}
\definecolor{eclipsePurple}{RGB}{127,0,85}
\definecolor{myred}{RGB}{100,0,0}
\definecolor{mygrey}{RGB}{211,211,211}
\definecolor{mygrey2}{RGB}{112,128,124}

\hypersetup{
	colorlinks = true,
	citecolor = {violet},
	linkcolor = {purple},
	urlcolor  = {blue}
}

\input{user_defined_commands.tex}

%% file: user_defined_commands.tex
\DeclareSymbolFont{symbolsC}{U}{txsyc}{m}{n}
\DeclareMathSymbol{\Searrow}{\mathrel}{symbolsC}{117}

\newcommand{\mathhl}[2]{\colorbox{#1}{$\displaystyle #2$}}

\newcommand{\lustre}{{\sc{Lustre}}}
\newcommand{\nlustre}{{\sc{NLustre}}}

\newcommand{\tbf}[1]{\textbf{#1}}
\newcommand{\ttt}[1]{\texttt{#1}}
\newcommand{\tti}[1]{\textit{#1}}

\newcommand{\type}[1]{\textcolor{eclipseBlue}{#1}}

\newcommand{\stlub}{\sqcup}

\newcommand{\rel}{\xspace \sqsubseteq }
\newcommand{\strel}{\xspace \type{\sqsubseteq} }
\newcommand{\lub}{\xspace \sqcup}

\newcommand{\for}[2]{#2\mapsto #1}

\newcommand{\val}[1]{\textcolor{mygrey2}{ #1 } }
\newcommand{\when}[2]{#1~\texttt{when}~#2} 
 
\newcommand{\binop}[2]{ #1 \oplus #2}
\newcommand{\unop}[1]{ \diamond ~#1}
\newcommand{\ite}[3]{\ttt{if } #1\ttt{ then } #2 \ttt{ else } #3}

\newcommand{\lmerge}[3]{\ttt{merge}~#1~#2~#3}
\newcommand{\ck}[2]{#1~\ttt{on}~#2}

\newcommand{\on}[3]{#1 ~\ttt{on}~ #2=\val{#3}}
\newcommand{\onF}[3]{#1 ~\ttt{on}~ #2=\val{\neg #3}}
\newcommand{\fby}[2]{#1~\texttt{fby}~#2}
\newcommand{\lfby}[2]{#1~\texttt{fby}_l~#2}
\newcommand{\nlfby}[2]{#1~\texttt{fby}_{nl}~#2}
\newcommand{\eqn}[3]{#1 =_{#3} #2}
\newcommand{\nodecall}[2]{#1 (\vv{#2})} 
\newcommand{\Node}{
  \ttt{node}~ f(\vv{x^{ck}}) ~ \texttt{returns} ~\vv{y^{ck}} \\
 {}&~\ttt{var}~ \vv{z} ~ \ttt{let}~ \vv{eqn} ~\ttt{tel}
}

\newcommand{\norm}[1]{\lfloor #1 \rfloor}

\newcommand{\node}{
 \Bigg\{ \begin{array}{c}
      \textsf{name} =~ \ttt{f};~ \textsf{in}=~\vv{x}; ~\textsf{var}=~\vv{z};\\
 \textsf{out}=~\vv{y};~ \textsf{eqs}=~\vv{eqn}\\
 \end{array}   \Bigg\}
}

\newcommand{\securitySignature}[5]{\stackrel{Node}{\vdash} \texttt{\textcolor{eclipsePurple}{Node} #1 } \textcolor{eclipseBlue}{#2^{#4}} \xrightarrow{#5} \textcolor{eclipseBlue}{#3}}

\newcommand{\vars}[2]{(#1_{1},\dots#1_{#2})}

\newcommand{\set}[1]{\{ #1\}}

\makeatletter
\newcommand{\leqnos}{\tagsleft@true\let\veqno\@@leqno}
\newcommand{\reqnos}{\tagsleft@false\let\veqno\@@eqno}
\reqnos
\makeatother

\newcommand{\mjdg}[2]{ \trfrac[]{ #1 }{  #2 } }
\newcommand{\namedJdg}[3]{ \trfrac[~(\footnotesize{#3})]{ #1 }{  #2 } }

\newcommand{\mexprPred}[3]{ #1 \textcolor{BrickRed}{\stackrel{e}{\vdash}} #2: \textcolor{eclipseBlue}{#3} }

\newcommand{\mcexprPred}[3]{ #1 \textcolor{OliveGreen}{\stackrel{ce}{\vdash}}  #2: \textcolor{eclipseBlue}{#3} }

\newcommand{\mclkPred}[3]{ #1 \textcolor{NavyBlue}{\stackrel{ck}{\vdash}}  #2: \textcolor{eclipseBlue}{#3} }

\newcommand{\meqnPred}[3]{  #1\stackrel{ eqn }{\vdash}  #2\ttt{~:>} ~{\type{#3}} }

\makeatletter
\newcommand{\predSet}[1]{%
    #1 \pchecknextarg}
\newcommand{\pchecknextarg}{\@ifnextchar\bgroup{\pgobblenextarg}{ }}
\newcommand{\pgobblenextarg}[1]{ ~~ #1 \@ifnextchar\bgroup{\pgobblenextarg}{ }}
\makeatother

\newcommand{\predSets}[2]{#1 \quad #2 }

\makeatletter

\newcommand{\dependSet}[1]{%
\begin{tralign}
#1  \checknextarg}
\newcommand{\checknextarg} {\@ifnextchar\bgroup {\gobblenextarg}{\end{tralign}}} 
\newcommand{\gobblenextarg}[1]{ \\[1.5mm] #1 \@ifnextchar\bgroup{\gobblenextarg} { \end{tralign}} }
\makeatother

\makeatletter
\newcommand{\union}[1]{%
    #1 \uchecknextarg}
\newcommand{\uchecknextarg}{\@ifnextchar\bgroup{\ugobblenextarg}{ }}
\newcommand{\ugobblenextarg}[1]{ \cup #1 \@ifnextchar\bgroup{\ugobblenextarg}{ }}
\makeatother

\makeatletter

\newcommand{\ichecknextarg}{\@ifnextchar\bgroup{\igobblenextarg}{ }}
\newcommand{\igobblenextarg}[1]{ \cap #1 \@ifnextchar\bgroup{\igobblenextarg}{ }}
\makeatother

\makeatletter

\newcommand{\orchecknextarg}{\@ifnextchar\bgroup{\orgobblenextarg}{ }}
\newcommand{\orgobblenextarg}[1]{ \bigvee #1 \@ifnextchar\bgroup{\orgobblenextarg}{ }}
\makeatother

\makeatletter

\newcommand{\andchecknextarg}{\@ifnextchar\bgroup{\andgobblenextarg}{ }}
\newcommand{\andgobblenextarg}[1]{ \land #1 \@ifnextchar\bgroup{\andgobblenextarg}{ }}
\makeatother

\newcommand*\substile[2]{%
  \,\scalebox{0.38}[0.5]{$\sststile[s]{\textstyle#1}{\textstyle#2}$}\,
}

\newcommand{\Proofsketch}{\textsc{Proof sketch.}\ \ }
\newcommand{\QED}{~\hfill $\square$}

\newcommand{\stream}[1]{\text{\guilsinglleft}\textcolor{mygrey2}{#1}\text{\guilsinglright}}

\newcommand{\nullStream}{\text{\guilsinglleft}\text{\guilsinglright}}
\newcommand{\ckFont}[1]{\textcolor{blue}{\textsf{#1}}}

\newcommand{\ignore}[1]{}

\newcommand{\streamSim}[3]{#1 \Downarrow_{#3} #2}

\newcommand{\expSim}[2]{\streamSim{#1}{#2}{\textsf{e}} }

\newcommand{\ckSim}[2]{\streamSim{#1}{#2}{\textsf{ck}}}

\newcommand{\flatten}[1]{\textcolor{blue}{\flat} (#1) }

\newcommand{\semConst}[3]{\textcolor{blue}{\textsf{const}}~#1~\val{#2}=
{\textcolor{BrickRed}{#3}}}

\newcommand{\whenk}[4]{\textcolor{blue}{\textsf{when}}~\val{#1}~{\textcolor{BrickRed}{#2}}~{\textcolor{BrickRed}{#3}}={\textcolor{BrickRed}{#4}}}

\newcommand{\mapwhenk}[4]{\textcolor{blue}{\widehat{\textsf{when}}}~\val{#1}~{\textcolor{BrickRed}{#2}}~{\textcolor{BrickRed}{#3}}={\textcolor{BrickRed}{#4}}}

\newcommand{\liftunop}[2]{\textcolor{blue}{\widehat{\diamond}} ~
{\textcolor{BrickRed}{#1}} = {\textcolor{BrickRed}{#2} }}

\newcommand{\liftbinop}[3]{{\textcolor{BrickRed}{#1}} \textcolor{blue}{\widehat{\oplus}} {\textcolor{BrickRed}{#2}} = {\textcolor{BrickRed}{#3}} }

\newcommand{\liftnode}[1]{ \textcolor{blue}{\widehat{#1}} }
\newcommand{\htl}[1]{(\textcolor{blue}{\textsf{htl}} ~#1)}
\newcommand{\tl}[1]{(\textcolor{blue}{\textsf{tl}} ~#1)}
\newcommand{\cc}[2]{(#1 \cdot #2)}
\newcommand{\ccnb}[2]{#1 \cdot #2}
\newcommand{\true}{\textcolor{blue}{\textsf{true}}}
\newcommand{\false}{\textcolor{blue}{\textsf{false}}}

\newcommand{\semFby}[3]{\textcolor{blue}{\textsf{fby}_{NL}} ~\val{#1} 
~{\textcolor{BrickRed}{#2}} = {\textcolor{BrickRed}{#3}}}

\newcommand{\semLFby}[3]{{\textcolor{blue}{\textsf{fby}_L}} ~{\textcolor{BrickRed}{#1}} ~{\textcolor{BrickRed}{#2}} = {\textcolor{BrickRed}{#3}}}

\newcommand{\mapsemLFby}[3]{{\textcolor{blue}{\widehat{\textsf{fby}_L}}} ~{\textcolor{BrickRed}{#1}} ~{\textcolor{BrickRed}{#2}} = {\textcolor{BrickRed}{#3}}}

\newcommand{\semLDFby}[4]{\textcolor{blue}{\textsf{fby$_{dl}$}} ~\val{#1} ~{\textcolor{BrickRed}{#2}} ~{\textcolor{BrickRed}{#3}} = {\textcolor{BrickRed}{#4}}}

\newcommand{\semMerge}[4]{\textcolor{blue}{\textsf{merge}}~{\textcolor{BrickRed}{#1}}~{\textcolor{BrickRed}{#2}}~{\textcolor{BrickRed}{#3}}={\textcolor{BrickRed}{#4}}}

\newcommand{\mapsemMerge}[4]{\textcolor{blue}{\widehat{\textsf{merge}}}~{\textcolor{BrickRed}{#1}}~{\textcolor{BrickRed}{#2}}~{\textcolor{BrickRed}{#3}}={\textcolor{BrickRed}{#4}}}

\newcommand{\semIte}[4]{\textcolor{blue}{\textsf{ite}}~{\textcolor{BrickRed}{#1}}~{\textcolor{BrickRed}{#2}}~{\textcolor{BrickRed}{#3}}={\textcolor{BrickRed}{#4}}}

\newcommand{\mapsemIte}[4]{\textcolor{blue}{\widehat{\textsf{ite}}}~{\textcolor{BrickRed}{#1}}~{\textcolor{BrickRed}{#2}}~{\textcolor{BrickRed}{#3}}={\textcolor{BrickRed}{#4}}}

\newcommand{\baseOf}[1]{\textcolor{blue}{\textsf{base-of}}\,#1}

\newcommand{\resClk}[3]{\textcolor{blue}{\textsf{respects-clock}}\,#1\,#2\,#3}

\newcommand{\hist}[1]{H_{*}(#1)}
\newcommand{\Hst}{H_{*}}

\newcommand{\tupstrm}[1]{\textcolor{BrickRed}{\mathbf{#1}}}

\newcommand{\mstrSemExpPred}[4]{#1,#2 ~\textcolor{BrickRed}{\vdash} ~\expSim{#3}{\textcolor{BrickRed}{#4}} }

\newcommand{\mstrSemCkPred}[4]{ #1,#2 ~\textcolor{NavyBlue}{\vdash} ~\ckSim{#3}{#4} }

\newcommand{\mstrEqnPred}[4]{#1,#2,#3 ~\textcolor{Black}{\vdash}~ #4}

\newcommand{\mstrCallPred}[4]{#1 ~\textcolor{Black}{\substile{s}{}}~ 
{#2}({\textcolor{BrickRed}{#3}}) \Searrow {\textcolor{BrickRed}{#4}} }

  \newcommand{\simpl}[2]{\ckFont{simplify}~#1~\type{#2}}
  \newcommand{\nlsimpl}[2]{\ckFont{simplify$_{NL}$}~#1~\type{#2}}
  \newcommand{\lsimpl}[2]{\textcolor{blue}{\textsf{simplify$_{L}$}}~#1~\type{#2}}

\lstdefinelanguage{Lustre}
{
  morekeywords={
    node,
    when,
    merge,
    fby,
    if,
    then,
    else,
    returns,
    let,
    tel,
    int,
    bool,
    var
  },
  basicstyle=\ttfamily, 
  captionpos=b,
  sensitive=true, 
  morecomment=[l]{--}, 
  morecomment=[s]{/*}{*/}, 
  morestring=[b]", 
  commentstyle=\color{eclipseGreen}, 
  keywordstyle=\color{eclipsePurple}, 
  stringstyle=\color{eclipseBlue}, 
  mathescape = true,
}

%% file: intro.tex
\section{Introduction}
\label{SEC:Introduction}

The synchronous reactive data flow language \lustre\ \cite{Lustre87,LusMain}  is an expressive language with an elegant formal semantics.
Its underlying deterministic, clocked model makes it a versatile programming paradigm, with diverse applications such as distributed embedded controllers, numerical computations, and complex Scade 6 \cite{Scade6} safety-critical systems.
It is also equipped with a suite of tools, comprising:
(a) a certified compilation framework from the high-level model into lower-level imperative languages \cite{SNLustre2017,Lustre2020};
(b) model-checkers \cite{LusMC,Kind2} 
(c) simulation tools \cite{Lurette} for program development.

The development of a formally certified compiler from \lustre\ to an imperative language is the subject of active research \cite{SNLustre2017,Lustre2020}.
A critical intermediate step involves the translation from \lustre\ to a well-behaved sub-language called ``Normalised \lustre'' (\nlustre), presented in \cite{Lustre2020}.
A recent paper (in French) defines the normalisation transformations from \lustre\ to  \nlustre, and establishes formally that they are semantics-preserving with respect to the stream semantics \cite{Bourke-jfla2021} (see Theorems \ref{THM:JFLA1} and \ref{THM:JFLA2}).

Recently we proposed a Denning-style lattice-based secure information flow (SIF) type system for \nlustre, and proved its soundness by establishing that securely-typed programs are \textit{non-interfering} with respect to the co-inductive stream semantics \cite{PYS-memocode2020}.
The main ideas underlying the security type system are (i) that each stream is assigned (w.r.t. assumptions on variables) a \textit{symbolic} security type, 
(ii) equations induce constraints on security types of the defined variables and of the defining expressions, and 
(iii) the output streams from a node have security levels at least as high as those of the input streams on which they depend. 
The symbolic constraint-based formulations allows us to  \textit{infer} constraints that suffice to ensure security.
The rules are simple, intuitive and amenable to being incorporated into the mechanised certified compilation \cite{Leroy2009} already developed for \lustre\ \cite{Velus}. 
In this paper, we propose a similar secure-information-flow type system for unrestricted \lustre\ (\S\ref{sec:LSec-types}).
The main innovation is formulating symbolic constraint-based \textit{refinement} (sub)types. 
These are necessitated by the presence in \lustre\ of nested node calls (in \nlustre,  directed nesting is disallowed). 

The security type system is shown by reduction to be \textit{sound} with respect to \lustre's \textit{co-inductive} stream semantics.
While it is possible to do so directly by establishing that well-security-typed programs exhibit non-interference \cite{GoguenM82a} using exactly the approach in \cite{PYS-memocode2020}, here we do so via a sound \textit{compiler transformation}: 
We show that the semantics-preserving normalisation transformations  (\textrm{de-nesting} and \textrm{distribution}, and \textrm{explicit initialisation} of \texttt{fby}) from \lustre\ to  \nlustre\ proposed in \cite{Bourke-jfla2021} \textit{preserve security types} as well (Theorem \ref{THM:SecPreserve} in \S\ref{SEC:Normalisation}).
In particular, there is a strong correspondence at the level of \textit{node definitions}. 
The preservation of \textit{security signatures} of node definitions is established via Lemma \ref{LEM:Simplify} in \S\ref{sec:LSec-types}.
The main idea is to remove local variable types via a substitution procedure \type{\textsf{simplify}} (Figure \ref{FIG:simpl}), showing that this maintains satisfiability of type constraints.
Since these transformations  preserve operational behaviour \textit{as well as security types of nodes}, and since we have already established non-interference for the target language \nlustre\ \cite[Theorem 5]{PYS-memocode2020}, 
non-interference holds for source \lustre\ as well (Theorem \ref{THM:soundness} in \S\ref{SEC:noninterference}).

Although the paper is intimately dependent on results of earlier work, we have endeavoured to keep it self-contained.
The reader interested in the complete stream semantics of \lustre\ may refer to the appendices.

\paragraph{Related Work.} \ \ 
We mention only the immediately relevant work here; a fuller discussion on related work can be found in \cite{PYS-memocode2020}.
The formalisation of \lustre\ semantics and its certified compilation are discussed in detail in \cite{Auger2013,SNLustre2017,Lustre2020}.
The normalisation transformations examined here are proposed in \cite{Bourke-jfla2021}.
Our lattice-based SIF framework harks back to Denning's seminal work \cite{Denning:1976}.
The idea of type systems for SIF can be found in \textit{e.g.}, \cite{Volpano96}.
That work also expressed soundness of a SIF type system in terms of the notion of non-interference \cite{GoguenM82a}.
Our previous work \cite{PYS-memocode2020} adapted that framework to a declarative data flow setting, showing that it is possible to \textit{infer} minimal \textit{partial-ordering constraints} between \textit{symbolic types}.  
The idea of type-preservation under the rewriting of programs is commonplace in logic and proof systems (``subject reduction'').

\ignore{
\noindent{\bf Structure of the paper:} \ \ 
In \S\ref{SEC:Lustre}, we review the language \lustre\ and its semantics. 
Our presentation of the Stream semantics of \lustre\ is consistent with the CompCert encoding  on a github repository \cite{Velus} mentioned in \cite{Lustre2020}. 

Based on these rules, we propose a definition of security for \lustre\ programs.

The main results follow in \S\ref{SEC:Soundness}, where we 
show, using the \textit{simple security} lemma for expressions and the \textit{constraints} for equations and programs, that securely-typed \lustre\ programs exhibit \textit{non-interference} with respect to the Stream semantics.

The paper concludes with a discussion of the related work (\S\ref{SEC:related}) and directions for future work in \S\ref{SEC:Conclusion}.
Auxiliary definitions such as those used in the Stream semantics are presented in Appendix \ref{SEC:Appendix-Auxiliary}. 

}

%% file: LustreSyntax.tex
 \section{\lustre\ and \nlustre}\label{SEC:Lustre}
    
\begin{figure*}[ht]

\begin{minipage}{.45\textwidth}
\small
    \begin{align*}
    e  :={}& \tag{\tbf{expr}} \\
       | {}&~ c \tag{cnst} \\
       | {}&~ x \tag{var} \\
       | {}&~ \unop{e} \tag{unop} \\
       | {}&~ \binop{e}{e} \tag{binop} \\
       | {}&~ \when{\vv{e}}{x=k} \tag{whn}\\
       | {}&~ \lmerge{x}{\vv{e}}{\vv{e}} \tag{mrg}\\
       | {}&~ \ite{e}{\vv{e}}{\vv{e}}  \tag{ite}\\
       | {}&~ \fby{\vv{e}}{\vv{e}}        \tag{fby} \\
       | {}&~ f(\vv{e}) \tag{ncall}\\
   eq  :={}& \tag{\tbf{equation}} \\
      |{}& \vv{x} = \vv{e} \tag{eq} \\
      \\
      \\
    \end{align*}
    \caption{\lustre\ syntax}\label{FIG:LustreSyntax}
\end{minipage}%
\begin{minipage}{.55\textwidth}
   \small
       \begin{align*}
       e  :={}& \tag{\tbf{expr'}} \\
          | {}&~ c \tag{cnst} \\
          | {}&~ x \tag{var} \\
          | {}&~ \unop{e} \tag{unop} \\
          | {}&~ \binop{e}{e} \tag{binop} \\
          | {}&~\when{e}{x=k}  \tag{whn'}\\
      ce  :={}& \tag{\tbf{cntrl expr}} \\
          | {}&~ e \tag{expr'} \\
          | {}&~ \lmerge{x}{ce}{ce} \tag{mrg'} \\
          | {}&~ \ite{e}{ce}{ce} \tag{ite'} \\
      eq  :={}& \tag{\tbf{equation}} \\
          | {}&~ \eqn{x}{ce}{ck} \tag{eq'} \\
          | {}&~ \eqn{x}{\fby{c}{e}}{ck} \tag{fby'} \\
          | {}&~ \eqn{\vv{x}}{\nodecall{f}{e}}{ck} \tag{ncall'} \\
       \end{align*}
       \caption{\nlustre\ syntax}\label{FIG:NLustreSyntax}
   \end{minipage}

   \begin{minipage}{.55\textwidth}
      \begin{align*}
      ck  :={}& \tag{\tbf{clock}} \\
       | {}&~ \ttt{base} \tag{base} \\
       | {}&~ \ck{ck}{(x=k)} \tag{on} \\
      \end{align*}
   \end{minipage}%
    \begin{minipage}{.45\textwidth}
      \begin{align*}
      d  :={}& \tag{\tbf{node declr}} \\
      | ~{}& \Node \\
      G:= {}& \vv{d} \tag{\tbf{program}}
      \end{align*}
    \end{minipage}

    
    \caption{Common syntax of nodes and clocks}\label{FIG:ClockNodeSyntax}

\end{figure*}
A \lustre\ program describes a synchronous network with \textit{clocked} streams of data flowing between operators and \textit{nodes}.
A program consists of a set of \textit{node definitions}, each parameterised by clocked input and output flows.
A clock is a boolean stream -- either a \texttt{base} clock or one \textit{derived} from another clock when a variable takes a specific (boolean) value (\texttt{on} $x=k$, where $k \in \{ \ttt{T},\ttt{F} \}$).

Each node comprises a set of (possibly mutually recursive) \textit{equations}, which define local variables and output flows in terms of flow \textit{expressions}.
Such definitions are unique, and may appear in any order.
\lustre\ satisfies the \textit{definition} and \textit{substitution} principles, namely that the context does not determine the meaning of an expression and that referential transparency holds.
Nodes do not have free variables. 
Nodes cannot make recursive calls; therefore, the dependency order on nodes forms a DAG.
All expressions and equations can be \textit{annotated} with a clock, following a static analysis to determine clock dependencies.

Figures \ref{FIG:LustreSyntax}--\ref{FIG:ClockNodeSyntax} 
present the syntax of \lustre\ and \nlustre. 

\lustre\ expressions (Figure \ref{FIG:LustreSyntax}) include flows described by constants, variables, unary and binary operations on flows, as well as the flows obtained by sampling when a variable takes a particular boolean value (\texttt{when}), interpolation based on a boolean variable flow (\texttt{merge}), and conditional combinations of flows (\texttt{if\_then\_else}).
Of particular interest are flows involving guarded delays (\texttt{fby}) and those involving \textit{node calls}.

\nlustre\ is a sub-language into which \lustre\ can be translated, from which subsequent compilation is easier. 
The main differences between between \lustre\ and \nlustre\ are 
(i) the former supports \textit{lists} of flows (written $\vv{e}$) for conciseness, whereas in the latter all flows are single streams;
(ii) \nlustre\ requires that conditional and \texttt{merge} ``control'' expressions are not nested below unary and binary operators or sampling;
(iii) node call and delayed flows (\texttt{fby}) are treated as first-class expressions, whereas in \nlustre, they can appear only in the context of equations;
(iv) \lustre\ permits nested node calls, whereas there is no nesting in \nlustre;
(v) finally,  the first argument of \texttt{fby} expressions in \nlustre\; must be a \textit{constant}, to enable a well-defined initialisation that can be easily implemented.

The \textit{translation} from \lustre\ to \nlustre\  \cite{Bourke-jfla2021} involves \textit{distributing} constructs over the individual components of lists of expressions, and \textit{de-nesting} expressions by introducing fresh local variables (See \S\ref{SEC:Normalisation}).
The reader can see an example, adapted from \cite{Bourke-jfla2021}, of a \lustre\ program and its translation into \nlustre\ in Figure \ref{FIG:L2NLEx} (ignoring for the moment the \type{security type annotations} therein).


\subsection{Stream Semantics}
\label{SEC:LustreStreamSem}

\ignore{
\begin{figure*}
    $$
    \begin{array}{c}
        \namedJdg{\semConst{c}{bs}{cs}}
        {\mstrSemExpPred{G,\Hst}{bs}{c}{[cs]}}{LScnst} ~~~
        \namedJdg{\Hst(x)={\textcolor{BrickRed}{xs}}}
        {\mstrSemExpPred{G,\Hst}{bs}{x}{[xs]}}{LSvar} \\
        \\
        \namedJdg{\predSet{\mstrSemExpPred{G,\Hst}{bs}{e}{[es]}}
            {\liftunop{es}{os} }}
        {\mstrSemExpPred{G,\Hst}{bs}{\unop{e}}{[os]}}{LSunop} \\
        \\
        \namedJdg{\predSet{\mstrSemExpPred{G,\Hst}{bs}{e_1}{[es_1]}}
            {\mstrSemExpPred{G,\Hst}{bs}{e_2}{[es_2]}}
            {\liftbinop{es_1}{es_2}{os} }}
        {\mstrSemExpPred{G,\Hst}{bs}{\binop{e_1}{e_2}}{[os]}}{LSbinop} \\
        \\
        \namedJdg{\predSet{\forall i ~\mstrSemExpPred{G,\Hst}{bs}{e_i}{\tupstrm{es}_i}}
            {\hist{x}={\textcolor{BrickRed}{xs}}}
            {\forall i:~\mapwhenk{k}{xs}{\tupstrm{es}_i}{\tupstrm{os}_i}}}
        {\mstrSemExpPred{G,\Hst}{bs}{\when{\vv{e_i}}{x=k}}{
        \flatten{\vv{\tupstrm{os}_i}}}}{LSwhn}\\
        \\
        \namedJdg{\dependSet{\predSet{\mstrSemExpPred{G,\Hst}{bs}{e}{[es]}}
        {\forall i:~ ~\mstrSemExpPred{G,\Hst}{bs}{et_i}{\tupstrm{ets}_i}}
        }
        {
           \forall j:~  \mstrSemExpPred{G,\Hst}{bs}{ef_j}{\tupstrm{efs}_j} 
            ~~~~
            \mapsemIte{es}{(\flatten{\vv{\tupstrm{ets}_i}})}{
            (\flatten{\vv{\tupstrm{efs}_j}})}{\tupstrm{os}}
            }}
        {\mstrSemExpPred{G,\Hst}{bs}{\ite{e}{\vv{et_i}}{\vv{ef_j}}}{\tupstrm{os}}}{LSite} \\
        \\
       \namedJdg{\dependSet{\predSet{\hist{x}=xs}{
       \forall i: ~\mstrSemExpPred{G,\Hst}{bs}{et_i}{\tupstrm{ets}_i}
       }}
            {\forall j: ~
            \mstrSemExpPred{G,\Hst}{bs}{ef_j}{\tupstrm{efs}_j}
            ~~~~
             \mapsemMerge{xs}{(\flatten{\vv{\tupstrm{ets}_i}})}{
            (\flatten{\vv{\tupstrm{efs}_j}})}{\tupstrm{os}}}
            }
       {\mstrSemExpPred{G,\Hst}{bs}{\lmerge{x}{\vv{et_i}}{\vv{ef_j}}}{\tupstrm{os}}}{LSmrg} \\
       \\
       \namedJdg{\predSet{\forall i:~ ~\mstrSemExpPred{G,\Hst}{bs}{e0_i}{\tupstrm{e0s}_i}
      ~~~~ \forall j:~ 
       \mstrSemExpPred{G,\Hst}{bs}{e_j}{\tupstrm{es}_j}}
            { ~\mapsemLFby{(\flatten{\vv{\tupstrm{e0s}_i}})}{
            (\flatten{\vv{\tupstrm{es}_j}})}{\tupstrm{os}}}}
        {\mstrSemExpPred{G,\Hst}{bs}{\fby{\vv{e0_i}}{\vv{e_j}}}{\tupstrm{os}}}{LSfby} \\
   \\
        \namedJdg{\forall i \in [1,k] ~ \mstrSemExpPred{G,\Hst}{bs}{e_i}{\tupstrm{es}_i} ~~~
        [\hist{x_1}, \ldots, \hist{x_n}] = 
        \flatten{\vv{\tupstrm{es}_i}}
        }
        {\mstrEqnPred{G}{\Hst}{bs}{\vv{x_j}=\vv{e_i}}}{LSeq} \\
  \\
    \namedJdg{\dependSet{\predSet{\node \in G}
        {\hist{n.\tbf{in}} = \tupstrm{xs}}
        {\baseOf{\tupstrm{xs}} = bs}}
        {\predSet{\hist{n.\tbf{out}} = \tupstrm{ys}}
        {\forall eq \in \vv{eq}:~ \mstrEqnPred{G}{\Hst}{bs}{eq}}
         }
    }{\mstrCallPred{G}{\liftnode{f}}{\tupstrm{xs}}{\tupstrm{ys}}}{LSndef} \\
    \\
           \namedJdg{\predSet{\mstrSemExpPred{G,\Hst}{bs}{\vv{e_i}}{\tupstrm{xs}}}
            {\mstrCallPred{G}{\liftnode{f}}{\tupstrm{xs}}{\tupstrm{ys}}}}
        {\mstrSemExpPred{G,\Hst}{bs}{f(\vv{e_i})}{\tupstrm{ys}}}{LSncall}
    \end{array}
    $$
    \caption{Stream semantics of \lustre}
    \label{FIG:LusStrSemNodeEqn}
\end{figure*}
}

The semantics of \lustre\ and \nlustre\  programs are \textit{synchronous}:
Each variable and expression defines a data stream which pulses with respect to a \textit{clock}.
A clock is a stream of booleans (CompCert/Coq's $\true$ and $\false$ in V\'{e}lus).
A flow takes its $n^\tti{th}$ value on the $n^\tti{th}$ clock tick, \textit{i.e.},  some value, written $\stream{v}$, is present at instants when the clock value is \true, and none (written $\nullStream)$ when it is \false.
The \textit{temporal operators} \ttt{when}, \ttt{merge} and \ttt{fby} are used to express the complex clock-changing and clock-dependent behaviours of sampling, interpolation and delay respectively.


Formally the stream semantics is defined using predicates over the program graph $G$, a (co-inductive) stream \textit{history} ($\Hst: \tti{Ident} \rightarrow \tti{value}~\tti{Stream}$) that associates value streams to variables, and a clock $bs$ \cite{Lustre2020,PYS-memocode2020,Bourke-jfla2021}.
Semantic operations on (lists of) streams are written in \textcolor{blue}{blue \textsf{sans serif}} typeface.
Streams are written in \textcolor{BrickRed}{red}, with lists of streams usually written in \textbf{\textcolor{BrickRed}{bold face}}.
All these stream operators, defined co-inductively,  enforce the clocking regime, ensuring  the presence of a value when the clock is \ckFont{true}, and absence when \ckFont{false}.

The predicate $\mstrSemExpPred{G,\Hst}{bs}{e}{\tupstrm{es}}$ relates an \textit{expression} $e$ to a \textit{list} of streams, written $\tupstrm{es}$.
A list consisting of only a single stream $\textcolor{BrickRed}{es}$ is explicitly denoted as $\textcolor{BrickRed}{[es]}$.
The semantics of \textit{equations} are expressed using the predicate 
$\mstrEqnPred{G}{\Hst}{bs}{\vv{eq_i}}$, which requires \textit{consistency} between the assumed and defined stream histories in $\Hst$ for the program variables, as induced by the equations.
Finally, the semantics of \textit{nodes} is given as a stream history transformer predicate
$\mstrCallPred{G}{\liftnode{f}}{\tupstrm{xs}}{\tupstrm{ys}}.$

We discuss here only some constructs which relate to the normalisation transformations.
Appendices \ref{APP:LustreStreamSem} and \ref{APP:AuxPred} present a complete account of the stream semantics for \lustre\ and \nlustre, consistent with \cite{Velus}.

\ignore{
Figure \ref{FIG:LusStrSemNodeEqn} presents the stream semantics for \lustre.
While rules for \textit{some} constructs have been variously presented \cite{SNLustre2017,Lustre2020,PYS-memocode2020,Bourke-jfla2021}, our presentation can be considered as a definitive consolidated specification of the operational semantics of \lustre, consistent with the V\'{e}lus compiler encoding \cite{Velus}.
}
\ignore{
Rule (LScnst) states that a constant $c$ denotes a constant stream of the value $\stream{c}$ pulsed according to given clock $bs$.  
This is effected by the semantic operator \ckFont{const}.
}
\[
        \namedJdg{\Hst(x)={\textcolor{BrickRed}{xs}}}
        {\mstrSemExpPred{G,\Hst}{bs}{x}{[xs]}}{LSvar}
\]
Rule (LSvar) associates a variable $x$ to the stream given by $\Hst(x)$. 
\ignore{
In rule (LSunop), \ckFont{$\hat{\diamond}$} denotes the operation $\diamond$ \textit{lifted} to apply instant-wise to the stream denoted by expression $e$.
Likewise in rule (LSbinop), the  binary operation $\oplus$ is applied paired point-wise to the streams denoted by the two sub-expressions (which should both pulse according to the same clock).
In all these rules, an expression is associated with a \textit{single} stream. 
\[
                \namedJdg{\predSet{\forall i ~\mstrSemExpPred{G,\Hst}{bs}{e_i}{\tupstrm{es}_i}}
            {\hist{x}={\textcolor{BrickRed}{xs}}}
            {\forall i:~\mapwhenk{k}{xs}{\tupstrm{es}_i}{\tupstrm{os}_i}}}
        {\mstrSemExpPred{G,\Hst}{bs}{\when{\vv{e_i}}{x=k}}{
        \flatten{\vv{\tupstrm{os}_i}}}}{LSwhn}
\]
 
The rule (LSwhn) describes \textit{sampling} whenever a variable $x$ takes the boolean value $k$, from the flows arising from a list of expressions $\vv{e_i}$,  returning a list of streams of such sampled values.
The predicate $\widehat{\ckFont{when}}$ \textit{maps} the predicate $\ckFont{when}$ to act on the corresponding components of \textit{lists} of streams, \textit{i.e.},  \[ 
\mapwhenk{k}{xs}{[es_1, \ldots, es_k]}{[os_1, \ldots, os_k]} 
~\textrm{abbreviates} ~ 
\bigwedge_{i \in [1,k]}~ \whenk{k}{xs}{es_i}{os_i}.
\]
(Similarly for the predicates
$\widehat{\ckFont{merge}}$, $\widehat{\ckFont{ite}}$, and $\widehat{\ckFont{fby}_L}$.  
The operation $\flatten{\_}$ flattens a list of lists (of possibly different lengths) into a single list. 
Flattening is required since expression $e_i$ may in general denote a \textit{list} of streams \textcolor{BrickRed}{$\tupstrm{es}_i$}.
}
\ignore{
The expression $\lmerge{x}{\vv{et}_i}{\vv{ef}_j}$ achieves (lists of) streams on a faster clock.
The semantics in rule (LSmrg) assume that for each pair of corresponding component streams from
$\flatten{\tupstrm{ets}_i}$ and
$\flatten{\tupstrm{efs}_j}$, a value is present in the first stream and absent in the second at those instances where $x$ has a true value $\stream{T}$, and complementarily, a value is present in the second stream and absent in the first when $x$ has a false value $\stream{F}$.
Both values must be absent when $x$’s value is absent.
These conditions are enforced by the auxiliary semantic operation \ckFont{merge}. 
In contrast, the conditional expression $\ite{e}{\vv{et}}{\vv{ef}}$
requires that all three argument streams $es$, and the corresponding components from $\flatten{\vv{\tupstrm{ets}_i}}$ and
$\flatten{\vv{\tupstrm{efs}_j}}$ pulse to the same clock.
Again, values are selected from the first or second component streams depending on whether the stream $es$ has the value $\stream{T}$ or $\stream{F}$ at a particular instant.
These conditions are enforced by the auxiliary semantic operation \ckFont{ite}. 
}
\[
       \namedJdg{\dependSet{\predSet{\forall i:~ ~\mstrSemExpPred{G,\Hst}{bs}{e0_i}{\tupstrm{e0s}_i}
      ~~~~ \forall j:~ 
       \mstrSemExpPred{G,\Hst}{bs}{e_j}{\tupstrm{es}_j}}}
       {\predSet{
             ~\mapsemLFby{(\flatten{\vv{\tupstrm{e0s}_i}})}{
            (\flatten{\vv{\tupstrm{es}_j}})}{\tupstrm{os}}}}
            }
        {\mstrSemExpPred{G,\Hst}{bs}{\fby{\vv{e0_i}}{\vv{e_j}}}{\tupstrm{os}}}{LSfby}
\]
A delay operation is implemented by $\fby{e0}{e}$.
The rule (LSfby) is to be read as follows.
Let each expression $e0_i$ denote a list of streams
$\tupstrm{e0s_i}$, and each expression $e_j$ denote a list of streams $\tupstrm{es_j}$.
The predicate $\widehat{\ckFont{fby}_L}$ \textit{maps} the predicate $\ckFont{fby}_L$ to act on the corresponding components of \textit{lists} of streams, \textit{i.e.},  \[ 
\mapsemLFby{\tupstrm{xs}}{\tupstrm{ys}}{\tupstrm{zs}} 
~\textrm{abbreviates} ~ 
\bigwedge_{i \in [1,m]}~ \semLFby{xs_i}{ys_i}{zs_i}
\]
(Similarly for the predicates
$\widehat{\ckFont{when}}$,
$\widehat{\ckFont{merge}}$, and $\widehat{\ckFont{ite}}$.) 
The operation $\flatten{\_}$ flattens a list of lists (of possibly different lengths) into a single list. 
Flattening is required since expression $e_i$ may in general denote a \textit{list} of streams \textcolor{BrickRed}{$\tupstrm{es}_i$}.
The output list of streams $\tupstrm{os}$ consists of streams whose first elements are taken from each stream in $\flatten{\vv{\tupstrm{e0s}_i}}$ with the rest taken from the corresponding component of $\flatten{\vv{\tupstrm{es}_j}}$.
%
\[
  \namedJdg{\forall i \in [1,..,k] ~ \mstrSemExpPred{G,\Hst}{bs}{e_i}{\tupstrm{es}_i} ~~~
        [\hist{x_1}, \ldots, \hist{x_n}] = 
        \flatten{\vv{\tupstrm{es}_i}}
        }
        {\mstrEqnPred{G}{\Hst}{bs}{\vv{x_j}=\vv{e_i}}}{LSeq}
\]
The rule (LSeq) for equations checks the consistency between the assumed meanings for the defined variables $x_j$ according to the history $\Hst$ with the corresponding components of the tuple of streams $\flatten{\vv{\tupstrm{es}_i}}$ to which a tuple of right-hand side expressions evaluates.

\[
            \namedJdg{\dependSet{\predSet{\node \in G}
        {~~\hist{g.\tbf{in}} = \tupstrm{xs}}
        }
        {\predSet{\hist{g.\tbf{out}} = \tupstrm{ys}}{\baseOf{\tupstrm{xs}} = bs}
        {\forall eq \in \vv{eq}:~ \mstrEqnPred{G}{\Hst}{bs}{eq}}
         }
    }{\mstrCallPred{G}{\liftnode{f}}{\tupstrm{xs}}{\tupstrm{ys}}}{LSndef}
\]
The rule (LSndef) presents the meaning given to the definition named $f$ of a node $g \in G$  as a stream list transformer. 
If history $\Hst$ assigns lists of streams to the input and output variables for a node in a manner such that the semantics of the equations in the node are satisfied, then the semantic function $\liftnode{f}$ transforms input stream list $\tupstrm{xs}$ to output stream list $\tupstrm{ys}$.
The operation \ckFont{base-of} finds an appropriate base clock with respect to which a given list of value streams pulse.
\[
       \namedJdg{\predSet{\mstrSemExpPred{G,\Hst}{bs}{\vv{e_i}}{\tupstrm{xs}}}
       {\mstrCallPred{G}{\liftnode{f}}{\tupstrm{xs}}{\tupstrm{ys}}}}
        {\mstrSemExpPred{G,\Hst}{bs}{f(\vv{e_i})}{\tupstrm{ys}}}{LSncall}
\]
The rule (LSncall) applies the stream transformer semantic function $\liftnode{f}$ to the stream list $\tupstrm{xs}$ corresponding to the tuple of arguments $\vv{e_i}$, and returns the stream list $\tupstrm{ys}$.

\ignore{
\paragraph{Clocks and clock-annotated expressions.} \ 
We next present rules for clocks.
Further, we  assume that all (\nlustre) expressions in equations can be clock-annotated, and present the corresponding rules.

\begin{figure*}
$$
   \begin{array}{c}
        \namedJdg{}
        {\mstrSemCkPred{\Hst}{bs}{\ttt{base}}{bs}}{LSbase}  \\ \\
        \namedJdg{
            \dependSet{\predSets{\mstrSemCkPred{\Hst}{bs}{ck}{\cc{\true}{bk}}}
            {\hist{x}=\cc{\stream{k}}{xs}}}
            {\mstrSemCkPred{\htl{\Hst}}{\tl{bs}}{\on{ck}{x}{k}}{bs'}}}
        {\mstrSemCkPred{\Hst}{bs}{\on{ck}{x}{k}}{\cc{\true}{bs'}}}{LSonT} \\

        \namedJdg{
            \dependSet{\predSets{\mstrSemCkPred{\Hst}{bs}{ck}{\cc{\false}{bk}}}
            {\hist{x}=\cc{\nullStream}{xs}}}
            {\mstrSemCkPred{\htl{\Hst}}{\tl{bs}}{\on{ck}{x}{k}}{bs'}}}
        {\mstrSemCkPred{\Hst}{bs}{\on{ck}{x}{k}}{\cc{\false}{bs'}}}{LSonA1} 
        \\
               \namedJdg{
            \dependSet{
             \predSet{\mstrSemCkPred{\Hst}{bs}{ck}{\cc{\true}{bk}}}
                {\hist{x}=\cc{\stream{k}}{xs}}
            }
            {\mstrSemCkPred{\htl{\Hst}}{\tl{bs}}{\onF{ck}{x}{k}}{bs'}}}
        {\mstrSemCkPred{\Hst}{bs}{\onF{ck}{x}{k}}{\cc{\false}{bs'}}}{LSonA2} \\ \\

        \namedJdg{\dependSet{\mstrSemExpPred{\Hst}{bs}{e}{[\ccnb{\nullStream}{es}]}}
        {\mstrSemCkPred{\Hst}{bs}{ck}{\ccnb{\false}{cs}}}}
    {\mstrSemExpPred{\Hst}{bs}{e::ck}{[\ccnb{\nullStream}{es}]}}{NSaeA} ~~
    
        \namedJdg{\dependSet{\mstrSemExpPred{\Hst}{bs}{e}{[\ccnb{\stream{v}}{es}]}}
            {\mstrSemCkPred{\Hst}{bs}{ck}{\ccnb{\true}{cs}}}}
        {\mstrSemExpPred{\Hst}{bs}{e::ck}{[\ccnb{\stream{v}}{es}]}}{NSae} 
    \end{array}
$$   
    \caption{Stream semantics of clocks and annotated expressions}
    \label{FIG:StrSemCk}
\end{figure*}

The predicate 
$\mstrSemCkPred{\Hst}{bs}{ck}{bs'}$ in \autoref{FIG:StrSemCk} defines the meaning of a \nlustre\ clock expression $ck$ with respect to a given history $\Hst$ and a clock $bs$ to be the resultant clock $bs'$.
The \ttt{on} construct lets us define 
coarser clocks derived from  a given clock --- whenever a variable $x$ has the desired value $k$ and the given clock is true. 
The rules (LSonT), (LSonA1), and (LSonA2) 
present the three cases: respectively when variable $x$ has the desired value $k$ and clock is true; the clock is false; and  the program variable $x$ has the complementary value and the clock is true.
The auxiliary operations \ckFont{tl} and \ckFont{htl},  give the tail of a stream, and the tails of streams for each variable according to a given history $\Hst$. 
Rules (NSaeA)-(NSae) describe the semantics of  clock-annotated expressions, where the output stream carries a value exactly when the clock is true.
}

\paragraph{Stream semantics for \nlustre.} \ \ 
The semantic relations for \nlustre\ are either identical to (as in constants, variables, unary and binary operations) or else the singleton cases of the rules for \lustre\ (as in \ttt{merge}, \ttt{ite}, \ttt{when}).
The main differences are in the occurrence of \ttt{fby} (now in a restricted form) and node call only in the context of equations, (which are clock-annotated).
\begin{figure*}
   $$
    \begin{array}{c}
       \namedJdg{\predSet{\mstrSemExpPred{\Hst}{bs}{e::ck}{[vs]}}
            {\semFby{c}{vs}{\hist{x}}}}
        {\mstrEqnPred{G}{\Hst}{bs}{\eqn{x}{\fby{c}{e}}{ck}}}{NSfby'} \\
\\  
    \namedJdg{\dependSet{\predSet{\node \in G}
        {\hist{n.\tbf{in}} = \tupstrm{xs}}
        {\baseOf{\tupstrm{xs}} = bs}}
        {\predSet{\resClk{\Hst}{bs}}
        {\hist{n.\tbf{out}} = \tupstrm{ys}}
        {\forall eq \in \vv{eq}:~ \mstrEqnPred{G}{\Hst}{bs}{eq}}
         }
    }{\mstrCallPred{G}{\liftnode{f}}{\tupstrm{xs}}{\tupstrm{ys}}}{NSndef'} \\
    \\
\ignore{
        \namedJdg{\mstrSemExpPred{\Hst}{bs}{e::ck}{\hist{x}}}
        {\mstrEqnPred{G}{\Hst}{bs}{\eqn{x}{e}{ck}}}{NSeq} 
\\ 
}
   
    \namedJdg{\predSet{\mstrSemExpPred{\Hst}{bs}{\vv{e}}{\tupstrm{vs}}}
    {\mstrSemCkPred{\Hst}{bs}{ck}{\baseOf{\tupstrm{vs}}}}
    {\mstrCallPred{G}{\liftnode{f}}{\tupstrm{vs}}{\vv{\hist{x_i}}}}}
            {\mstrEqnPred{G}{\Hst}{bs}
        {\eqn{\vv{x}}{f(\vv{e})}{ck}  }}{NSncall'}
   
   \end{array}%
    $$
    
    \caption{Stream semantics of \nlustre\ nodes and equations}
    \label{FIG:StrSemNodeEqn}
\end{figure*}

The (NSfby') rule for \ttt{fby} in an equational context uses the semantic operation \textcolor{blue}{$\textsf{fby}_{NL}$}, which differs from $\ckFont{fby}_L$ in that it requires its first argument to be a constant rather than a stream.
The (NSndef') rule only differs from (LSndef) in that  after clock alignment during \textit{transcription}, we have an additional requirement of $\Hst$ being in accordance with the base clock $bs$, enforced by \ckFont{respects-clock}.
\ignore{
The (NSeq) rule for simple equations mentions the clock that annotates the defining expression, checking that it is consistent with the assumed history for the defined variable $x$.
}
Finally, the rule rule (NSncall') for node call, now in an equational context, is similar to (LSncall) except that it constrains the clock modulating the equation to be the base clock of the input flows.

%% file: LustreSecTyping.tex
\section{A Security Type System for {\lustre}}\label{sec:LSec-types}

We define a symbolic secure information flow type system, where under security-level type assumptions for program variables, \lustre\ expressions are given a \textit{security type}, and \lustre\ equations induce a set of \textit{ordering constraints} over security types.

Security type expressions ($\type{\alpha, \beta}$) for \lustre\ are either (i) \textit{type variables}  (written $\type{\delta})$ drawn from a set \textit{STV}, or (ii) constructed using a \textit{join} operator (written $\type{\alpha \stlub \beta}$). 
(iii) The identity element of the associative, commutative and idempotent monoid operation $\type{\stlub}$ is $\type{\bot}$.
While the above suffice for \nlustre, for \lustre we introduce (iv) \textit{refinement types} $\type{\alpha \{\!| \rho |\!\} }$, where symbolic constraint $\type{\rho}$ modulates type expression $\type{\alpha}$.
Constraints on security types, typically $\type{\rho}$,  are (conjunctions of) relations of the form $\type{\alpha \strel \beta}$.
Since security types are to be interpreted with respect to a complete lattice, we have $\type{\alpha \strel \beta}$ exactly when
$\type{\alpha \stlub \beta ~=~ \beta}$.
Our proposed security types and their equational theory are presented in Figure \ref{FIG:SecurityTypes}. 
The security types for \nlustre\  and their equational theory \cite{PYS-memocode2020} are highlighted in grey within the diagram.  
This congruence on \nlustre\ types (henceforth $\equiv_{NL}$), which is given in the highlighted second line of Figure \ref{FIG:SecurityTypes}, is significantly simpler since it does not involve refinement types!

\begin{figure*}
        \text{Types: }  
        $\type{\alpha,\beta,\gamma,\theta} ~::= \mathhl{mygrey}{\type{\bot}}  \;|\; \mathhl{mygrey}{\type{\delta} \in \textit{STV}}  \;|\; 
        \mathhl{mygrey}{\type{\alpha \lub \beta}} \;|\; \type{\alpha\{\!| \rho |\!\} }$  ~~
        \text{Constraints:} $\mathhl{mygrey}{\type{\rho} ~::=~ \type{(\theta \rel \alpha)}^*}$ \\ \\
$\mathhl{mygrey}{\type{(\alpha \lub \beta) \lub \theta} ~=~ \type{\alpha \lub (\beta \lub \theta)}, 
~\hfill~
\type{\alpha \lub \alpha} ~=~ \type{\alpha},
~\hfill~
\type{\alpha \lub \beta} ~=~ \type{\beta \lub \alpha},
~\hfill~
\type{\alpha \lub \bot} ~=~ \type{\alpha} ~=~ 
\type{\bot \lub \alpha},}$ \\ \\
$ \type{\alpha\{\!| |\!\}} ~=~ \type{\alpha},     
~~\hfill~~
\type{\alpha_1\{\!| \rho_1 |\!\}} \type{~\stlub}~ \type{\alpha_2\{\!| \rho_2 |\!\}} ~=~ (\type{\alpha_1 \lub \alpha_2}) \type{\{\!|} \union{ \type{\rho_1}} {\type{\rho_2}} \type{|\!\}}, 
~~\hfill~~
\type{\alpha\{\!| \rho_1 |\!\}\{\!| \rho_2 |\!\}} ~=~ \type{\alpha \{\!| } \union{\type{\rho_1}}{\type{\rho_2}} \type{|\!\}}$, \\  \\
$\type{\vv{\alpha_i} \{\!| \rho |\!\}} ~=~ 
\type{ \vv{ \alpha_i  \{\!| \rho |\!\}} }
$,
~~
$\{ \type{\alpha\{\!| \rho_1 |\!\} \rel \beta\{\!| \rho_2 |\!\}} \} ~=~ \union{\{ \type{\alpha \strel \beta} \} } {\union{\type{\rho_1}}{\type{\rho_2}}}$,
~~~
$\mathhl{mygrey}{\type{\vv{\alpha_j}[\theta_i/\delta_i]} ~=~ \type{\vv{\alpha_j[\theta_i/\delta_i]}}}$,
\\ \\
%
$\type{\alpha\{\!| \rho |\!\}[\theta_i/\delta_i]} ~=~ \type{\alpha[\theta_i/\delta_i] \{\!| \rho[\theta_i/\delta_i] |\!\}}$,
~~\hfill~~
$\mathhl{mygrey}{(\type{\alpha \strel \beta}) \type{[\theta_i/\delta_i]} ~=~
\type{\alpha[\theta_i/\delta_i] ~\strel~ \beta[\theta_i/\delta_i]}} 
$.

    \caption{Security types, constraints and their properties}
    \label{FIG:SecurityTypes}
\end{figure*}

We write  $\type{\alpha [ \theta_i / \delta_i]}$ for $i=1, \ldots, k$ to denote the (simultaneous) substitution of security types $\type{\theta_i}$ for security type variables $\type{\delta_i}$ in security type $\type{\alpha}$. 
The notation extends to substitutions on tuples ($\vv{\type{\alpha}}$) and constraints ($\type{\alpha \strel \beta}$).

Constraints are interpreted in a security class lattice $\tti{SC}$ by the homomorphic extension of a ground instantiation $s: \tti{STV} \rightarrow \tti{SC}$, such that 
$s(\type{\bot}) = \bot$,
$s(\type{\alpha \stlub \beta}) = s(\type{\alpha}) \lub s(\type{\beta})$,
$s(\type{\vv{\alpha}}) = \vv{s(\type{\alpha_i})}$,
$s(\type{\alpha \strel \beta}) = 
s(\type{\alpha}) \rel s(\type{\beta})$, and
$s(\type{\alpha\{\!| \rho |\!\}}) = s(\type{\alpha})$ if $s(\type{\rho})$ holds in $SC$, \textit{i.e.}, if ``$s$ satisfies $\type{\rho}$'' (else undefined).
To make sense, refinement type $\type{\alpha \{\!| \rho |\!\}}$ requires the satisfaction of constraint $\type{\rho}$.

\begin{lemma}[Soundness]
The equational theory induced by the equalities in Figure \ref{FIG:SecurityTypes} is sound with respect to any ground instantiation $s$, \textit{i.e.}, 
(i) $\type{\alpha = \beta}$ implies $s(\type{\alpha}) = s(\type{\beta})$, and (ii) $\type{\rho_1} =  \type{\rho_2}$ implies $s(\type{\rho_1})$ is satisfied iff $s(\type{\rho_2})$ is.
\end{lemma}

\begin{lemma}[Confluence]
All equations other than those of associativity and commutativity (AC) can be oriented left-to-right into rewriting rules.
The rewriting system is confluent modulo AC.
Equal types (respectively, equal constraints) can be rewritten to a common form modulo AC.
\end{lemma}
\Proofsketch
The equational theory $\equiv_{NL}$ is decidable, since it is a convergent rewriting system modulo AC.
The rules in lines 3 and 4 can all be oriented left to right. 
We use completion \cite{Knuth-Bendix} to introduce rule
$\type{\alpha_1\{\!| \rho_1 |\!\}} \type{~\stlub}~ \type{\alpha_2} \longrightarrow (\type{\alpha_1 \lub \alpha_2}) \type{\{\!|}  \type{\rho_1}  \type{|\!\}}$, when $\type{\alpha_2}$ is not a refinement type.
We use the theory of strongly coherent rewriting modulo AC \cite{Viry-rewriting}, to efficiently decide type equality. 
\QED.

\subsection{Security Typing Rules}
Assume typing environment $\Gamma: \tti{Ident} \rightarrow \tti{ST}$, a partial function associating a security type to each free variable \tti{x} in a \lustre\ program phrase.
Expressions and clocks are type-checked with the predicates:
$
\mexprPred{\Gamma}{e}{\vv{\alpha}}$
and 
$\mclkPred{\Gamma}{ck}{\alpha}$ respectively. 
These are read as ``under the context $\Gamma$ mapping variables to security types, $e$, $ck$ have security type(s) $\type{\alpha}$''.
The types for tupled expressions are tuples of the types of the component expressions.
For equations, we use the predicate:
$\meqnPred{\Gamma}{eq}{\rho}$, 
which states that under the context $\Gamma$, equation $eq$ when type-elaborated generates constraints $\type{\rho}$.
Elementary constraints for equations are of the form $\type{\alpha \rel \beta}$, where $\type{\beta}$ is the security type of the defined variable, and $\type{\alpha}$ the security type obtained from that of the defining expression joined with the clock's security type.
Since every flow in \lustre\ is defined \textit{exactly once}, by the Definition Principle, no further security constraints apply.

The security typing rules for \lustre\ are presented in Figures \ref{FIG:LustreClkSecTyping} -- \ref{FIG:LustreEqnType}, plus the rules for node definition and node call.
These rules generalise those in \cite{PYS-memocode2020} to handle expressions representing lists of flows, and nested node calls.
The rules for \nlustre\ expressions other than node call and \ttt{fby} are just the singleton cases. 
Node call and \ttt{fby} are handled by the rule for equations. 
\begin{figure*}
    \[
    \begin{array}{c}
        \namedJdg{\Gamma(\ttt{base}) = \type{\gamma}}{\mclkPred{\Gamma}{\ttt{base}}{\gamma}}{LTbase} ~~~~~~
        \namedJdg{\predSet{\Gamma(x)=\type{\gamma_1}}{\mclkPred{\Gamma}{ck}{\gamma_2}}}
        {\mclkPred{\Gamma}{\ck{ck}{x=k}}{\gamma_1 \stlub \gamma_2}}{LTon} \\
    \end{array}
    \]
    \caption{\lustre\ security typing rules for clocks}
\label{FIG:LustreClkSecTyping}
 \end{figure*}

 \begin{figure*}
    \[
    \begin{array}{c}
        \namedJdg{\predSet{\Gamma(x)=\type{\alpha}}}
        {\mexprPred{\Gamma}{x}{\alpha }}{LTvar} 
        ~~~~~~
        \namedJdg{\predSet{\mexprPred{\Gamma}{e }{\alpha}}}
        {\mexprPred{\Gamma}{{\unop{e}}}{\alpha}}{LTunop}  
~~~~~~
        \namedJdg{
            \predSet{\mexprPred{\Gamma}{e_1}{\alpha_1}} 
            {\mexprPred{\Gamma}{e_2}{\alpha_2}}}
        {\mexprPred{\Gamma}{\binop{e_1}{e_2}}{\alpha_1 \stlub \alpha_2 }} {LTbinop} \\
        \namedJdg{
            \predSet{\type{\theta}=\Gamma(x)}
            {\mexprPred{\Gamma}{\vv{e_t}}{\vv{\alpha}}}
            {\mexprPred{\Gamma}{\vv{e_f}}{\vv{\beta}}}}
    {\mexprPred{\Gamma}{ \lmerge{x}{\vv{e_t}}{\vv{e_f}}} {\vv{(\theta \stlub \alpha_i \stlub \beta_i)_i}}}{LTmrg} 
        ~~~~~
       \namedJdg{}
        {\mexprPred{\Gamma}{c}{\bot}} {LTcnst} \\
    
    \namedJdg{
            \predSet {\mexprPred{\Gamma}{e}{\theta}} 
            {\mexprPred{\Gamma}{\vv{e_t}}{\vv{\alpha}}} 
            {\mexprPred{\Gamma}{\vv{e_f}}{\vv{\beta}}} }
    {\mcexprPred{\Gamma}{\ite{e}{\vv{e_t}}{\vv{e_f}}}{ 
    \vv{(\theta \stlub \alpha_i \stlub \beta_i)_i}}}{LTite} \\
\\
    \namedJdg{
            \predSet{\mexprPred{\Gamma}{\vv{e_0}}{\vv{\alpha}}}
            {\mexprPred{\Gamma}{\vv{e}}{\vv{\beta}}}}
    {\mexprPred{\Gamma}{ \lfby{\vv{e_0}}{\vv{e}}} {\vv{(\alpha_i \stlub \beta_i)_i}}}{LTfby} 
    ~~
     \namedJdg{
          \predSet{\mexprPred{\Gamma}{e_1}{\alpha_1}} 
            \ldots
            {\mexprPred{\Gamma}{e_n}{\alpha_n}}~~
            {\Gamma(x)= \type{\gamma}}}
        {\mexprPred{\Gamma}{\when{\vv{e}}{x = k}}{ 
        \vv{(\alpha_i \stlub \gamma)_i}}}{LTwh}
    \end{array}
\]  
    \caption{\lustre\ Security Typing Rules for Expressions}
\label{FIG:LustreExpSecTyping}
\end{figure*}

 \begin{figure*}
\[
   \begin{array}{c}
    \namedJdg { 
         \predSet {\type{\vv{\beta}} = \Gamma(\vv{x})} 
         {\mexprPred{\Gamma}{\vv{e}}{\vv{\alpha}}}
         {\mclkPred{\Gamma}{ck}{\gamma}}}
     {\meqnPred{\Gamma}{ \vv{x}^{ck} = \vv{e}}{\type{\set{ (\gamma \stlub \alpha_i  \rel \beta_i)_i}}}} {LTeq} 
  \;
   \namedJdg{
        \predSets{\meqnPred{\Gamma}{eq}{\type{\rho}}} 
        {\meqnPred{\Gamma}{eqs}{\type{\rho'}}}}
   {\meqnPred{\Gamma}{eq;eqs}{\union{\type{\rho}}{\type{\rho'}}}}{LTeqs}
   \end{array}
 \]
   \caption{\lustre\ security typing rules for equations}
   \label{FIG:LustreEqnType}
\end{figure*}

In (LTbase), we assume $\Gamma$ maps the base clock $\ttt{base}$ to some security variable ($\type{\gamma}$ by convention).
In (LTon), the security type of the derived clock is the join of the security types of the clock $ck$ and that of the variable $x$.

Constants have security type $\type{\bot}$, irrespective of the context (rule (LTcnst)).
For variables, in rule (LTvar), we look up their security type in the context $\Gamma$.
Unary operations preserve the type of their arguments (rule (LTunop)).
Binary ($\oplus$,\ttt{when} and \ttt{fby}) and ternary (\ttt{if-then-else} and \ttt{merge}) operations on flows generate a flow with a security type that is the join of the types of the operand flows (rules (LTbinop), (LTwhn), (LTmrg), (LTite), and (LTfby).
In operations on \textit{lists of flows}, the security types are computed component-wise.
There is an implicit dependency on the security level of the common clock of the operand flows for these operators.
This dependence on the security level of the clock is made explicit in the rule for equations.
In general, the security type for any constructed expression is the join of those of its components (and of the clock).

\paragraph{Node call.}\ 
Node calls assume that we have a security signature for the node definition (described below).
We can then securely type node calls by instantiating the security signature with the types of the actual arguments (and that of the base clock).
Note the rule (LTncall) creates  refinement types consisting of the output types $\type{\beta_i}$ modulated with $\type{\rho'}$, the instantiated set of constraints $\type{\rho}$ taken from the node signature:
\[
\namedJdg{
 {\predSet
 { \securitySignature{f}{(\vv{\type{\alpha}})}{\vv{\type{\beta}}}{\type{\gamma}}{\type{\rho}}}~~
 {\mexprPred{\Gamma}{ \vv{e}} {\vv{\type{\alpha'}}}}~~
 {\Gamma(\ttt{base})=\type{\gamma'}}~~
 {\type{\rho'} = 
 \type{\rho[\gamma'/\gamma][\vv{\alpha'}/\vv{\alpha}]}}}}
 {\mexprPred{\Gamma}{ f(\vv{e})   }{\vv{\beta}\{\!| \type{\rho'} |\!\}}}{LTncall}
\]

\paragraph{Node definition.}\ 
A node definition is given a signature
$\securitySignature{f}{(\vv{\type{\alpha}})}{\vv{\type{\beta}}}{\type{\gamma}}{\type{\rho}}$, which is to be read as saying that the node named $f$ relates the security types $\vv{\type{\alpha}}$ of the input variables (and $\type{\gamma}$, that of the base clock) to the types of the output variables $\vv{\type{\beta}}$, via the constraints 
$\type{\rho}$.

Let $\type{\alpha_1,\dots,\alpha_n,\delta_1,\dots\delta_k,\beta_1,\dots\beta_m,\gamma}$ be distinct \textit{fresh type variables}.
Assume these to be the types of the input, local and output variables, and that of the base clock. 
We compute the constraints over these variables induced by the nodes equations.
Finally, we eliminate, via substitution, the type variables given to the local program variables, since these should not appear in the node's interface.
The security signature of a node definition is thus given as:
\[
\namedJdg{\dependSet{G(f)= n:\{\ttt{in}=\vv{x}, \ttt{out}=\vv{y}, \ttt{var}=\vv{z}, \ttt{eqn} = \vv{eq} \}}
    {\Gamma_F := \{ \for{\type{\vv{\alpha}}}{\vv{x}}, \for{\type{\vv{\beta}}}{\vv{y}}, \ttt{base} \mapsto \type{\gamma} \}~~~~
     \Gamma_L := \{\for{\vv{\type{\delta}}}{\vv{z}}\}}
    {\predSet{\meqnPred{\union{\Gamma_F}{\Gamma_L}}{\vv{eq}}{\type{\rho'}}}{(\_, \type{\rho}) = \simpl{(\_,\type{\rho'})}{\vv{\delta}}}}
    } 
    { \securitySignature{f}{(\vv{\type{\alpha}})}{\vv{\type{\beta}}}{\type{\gamma}}{\type{\rho}}}{LTndef}
\]
The node signature (and call) rules can be formulated in this step-wise and modular manner since \lustre\ does not allow recursive node calls and cyclic dependencies.
Further, all variables in a node definition are explicitly accounted for as input and  output parameters or local variables, so no extra contextual information is required. 

\begin{figure*}
\[
\begin{array}{cl}
    \mjdg{}
    {(\type{\vv{\alpha}, \rho}) = \simpl{(\type{\vv{\alpha}, \rho})}{[~]}} 
    ~~~~~
    \mjdg{(\type{\vv{\alpha'}, \rho'}) = \simpl{(\type{\vv{\alpha}}[\type{\nu/\delta}],
    \type{\rho}[\type{\nu/\delta}])}{\type{\vv{\delta}}}}
    {(\type{\vv{\alpha'}, \rho'}) = \simpl{(\type{\vv{\alpha}}, \union{\type{\rho}}{\{\type{\nu \rel \delta}\}})}{(\type{\delta} :: \type{\vv{\delta}})}}
     & \mbox{$\type{\delta}$ not in $\type{\nu}$}
\\ \\
    \mjdg{(\type{\vv{\alpha'}, \rho'}) = \simpl{(\type{\vv{\alpha}}[\type{\nu/\delta}], \type{\rho}[\type{\nu/\delta}])}{\type{\vv{\delta}}}}
    {(\type{\vv{\alpha'}, \rho'}) = \simpl{(\type{\vv{\alpha}}, \union{\type{\rho}}{\{\type{\nu \stlub \delta ~\rel~ \delta}\}})}{(\type{\delta} :: \type{\vv{\delta}})}}
    & \mbox{$\type{\delta}$ not in $\type{\nu}$}
\end{array}\]

\caption{Eliminating local variables' security type constraints}\label{FIG:simpl}
\end{figure*}

Observe that in the (LTndef) rule,  $\type{\delta_i}$ are fresh security type variables assigned to the local variables in a node. 
Since there will be exactly one defining equation for any local variable $z_i$, note that in constraints $\type{\rho'}$, there will be exactly one constraint in which $\type{\delta_i}$ is on the right, and this is of the form $\type{\nu_i \strel \delta_i}$.
Procedure \type{\textsf{simplify}} (Figure \ref{FIG:simpl}) serially (in some arbitrary but fixed order) eliminates such type variables via substitution in the types and type constraints. 
Our definition of \ckFont{simplify} here generalises that given for the types of \nlustre\ in \cite{PYS-memocode2020}.

\begin{lemma}[Correctness of $\simpl{(\type{\vv{\alpha},\rho})}{\type{\vv{\delta}}}$]\label{LEM:Simplify}
Let $\type{\rho}$ be a set of constraints such that for a security type variable $\type{\delta}$, there is at most one constraint of the form $\type{\mu \rel \delta}$.
Let $s$ be a ground instantiation of security type variables in a security class lattice $\tti{SC}$ such that $\type{\rho}$ is satisfied by $s$.
\begin{enumerate}
\item If $\type{\rho} = \union{\type{\rho_1}}{\set{\type{\nu \rel \delta}}}$, where variable $\type{\delta}$ is not in $\type{\nu}$,
then $\type{\rho_1}[\type{\nu}/\type{\delta}]$ is satisfied by $s$. (Assume disjoint union.)
\item If $\type{\rho} = \union{\type{\rho_1}}{\set{\type{\nu \stlub \delta \rel \delta}}}$, where variable $\type{\delta}$ is not in $\type{\nu}$,
then $\type{\rho_1}[\type{\nu}/\type{\delta}]$ is satisfied by $s$. (Assume disjoint union.)
\end{enumerate}
\end{lemma}

Lemma \ref{LEM:Simplify} is central to establishing that the type signature of a node does not change in the normalisation transformations of \S\ref{SEC:Normalisation}, which introduce equations involving fresh local program variables.

Revisiting Figure \ref{FIG:L2NLEx}, the reader can see the type system at work, with the \type{security types and constraints}
annotated. 
Also shown is the simplification of constraints using \ckFont{simplify}.

%% file: normalisation.tex
\section{Normalisation}
\label{SEC:Normalisation}

\begin{figure*}
 \[
 \begin{array}{rcl}
        \norm{c} &=&  ([c^{\type{\bot}}],[\;]^{\type{\emptyset}}) ~~\hfill \textit{Xcnst} \\
        \norm{x^{\type{\alpha}}} &=& 
                 ([x]^{\type{\alpha}}, [\;]^{\type{\emptyset}})
                  ~~\hfill \textit{Xvar} 
                  \\ \\
        \norm{\unop{e_1}} &=& \textrm{let}~ 
            ([e']^{\type{\alpha}},eqs^{\type{\rho}}) \leftarrow \norm{e}  ~~\hfill \textit{Xunop} \\
        && \textrm{in}~
        ([\unop{e'_1}]^{\type{\alpha}}, eqs^{\type{\rho}}) 
        \\ \\
        \norm{\binop{e_1}{e_2}} &=& \textrm{let}~
            ([e'_1]^{\type{\alpha_1}}, eqs_1^{\type{\rho_1}}) \leftarrow \norm{e_1} ~\textrm{and}~ ([e'_2]^{\type{\alpha_2}},eqs_2^{\type{\rho_2}}) \leftarrow \norm{e_2} ~~\hfill \textit{Xbinop} \\
        && \textrm{in}~
        ([\binop{e'_1}{e'_2}]^{\type{\alpha_1 \stlub \alpha_2}}, (\union{eqs_1}{eqs_2})^{\union{\type{\rho_1}}{\type{\rho_2}}}) 
        \\ \\
        \norm{\when{\vv{e}}{x^{\type{\gamma}}=k}} &=& \textrm{let}~
        ([e'_1{}^{\type{\alpha_1}}, \ldots, e'_m{}^{\type{\alpha_m}}], eqs^{\type{\rho}})  \leftarrow \norm{\vv{e}} ~~\hfill \textit{Xwhn}\\
        && \textrm{in}~ 
        ([\when{e'_1}{x=k}^{\type{\alpha_1 \stlub \gamma}}, \ldots,  \when{e'_m}{x=k}^{\type{\alpha_m \stlub \gamma}} ], eqs^{\type{\rho}}) 
        \\ \\
        \norm{\fby{\vv{e_0}}{\vv{e_1}}} &=& 
        \textrm{let}~
        (\vv{e'_0}{}^{\type{\vv{\alpha}}}, eqs_0^{\type{\rho_0}})   \leftarrow \norm{\vv{e_0}} ~\textrm{and}~
     (\vv{e'_1}{}^{\type{\vv{\beta}}}, eqs_1^{\type{\rho_1}})   \leftarrow \norm{\vv{e_1}}  ~~\hfill \textit{Xfby} \\
    && \textrm{in}~ 
    (\vv{x}^{\type{\vv{\delta}}},  
        (\{ ( x_i=\fby{e'_{0i}}{e'_{1i}} )_{i=1}^k
                 \} \cup {eqs_0} \cup {eqs_1})^{\type{\rho}}) \\
    &&     \textrm{where}~     \type{\rho} = \{ ( \type{\alpha_i \stlub \beta_i \strel \delta_i} )_{i=1}^k \} \cup
    \type{\rho_0} \cup \type{\rho_1}
                 \\ \\
    \norm{\lmerge{x^{\type{\gamma}}}{\vv{e_1}}{\vv{e_2}}} &=& \textrm{let}~ (\vv{e'_1}^{\type{\vv{\alpha}}},eqs_1^{\type{\rho_1}}) \leftarrow \norm{\vv{e_1}} ~\textrm{and}~  (\vv{e'_2}^{\type{\vv{\beta}}},eqs_2^{\type{\rho_2}}) \leftarrow \norm{\vv{e_2}} ~~\hfill \tti{Xmrg} \\
       && \textrm{in}~
       (\vv{x}^{\type{\vv{\delta}}},  
        (\{ ( x_i=\lmerge{x}{e'_{1i}}{e'_{2i}} )_{i=1}^k
                 \} \cup {eqs_1} \cup {eqs_2})^{\type{\rho}}) \\
    &&     \textrm{where}~     \type{\rho} = \{ ( \type{\gamma \stlub \alpha_i \stlub \beta_i \strel \delta_i} )_{i=1}^k \} \cup
    \type{\rho_1} \cup \type{\rho_2} \\ \\
    \lfloor\ttt{if}~ e~ \ttt{then}~ \vv{e_t}
    \\ 
     ~\ttt{else} ~{\vv{e_f}}\rfloor &=& 
     \textrm{let}~ (e'^{\type{\kappa}},eqs_c^{\type{\rho_c}}) \leftarrow \norm{e} ~\textrm{and}~ (\vv{e'_t}^{\type{\vv{\alpha}}},eqs_t^{\type{\rho_t}}) \leftarrow \norm{\vv{e_t}} ~~\hfill \tti{Xite} \\ 
    &&\textrm{and}~  (\vv{e'_f}^{\type{\vv{\beta}}},eqs_f^{\type{\rho_f}}) \leftarrow \norm{\vv{e_f}} ~ \textrm{in}~ \\
    && 
       (\vv{x}^{\type{\vv{\delta}}},  
        (\{ ( x_i=\ite{e'}{e'_{ti}}{e'_{fi}} )_{i=1}^k
                 \} \cup {eqs})^{\type{\rho}}) \\
    && \textrm{where}~ eqs = \union{eqs_c}{eqs_t}{eqs_f} \\
    &&\type{\rho} = \union{( \type{\kappa \stlub \alpha_i \stlub \beta_i \strel \delta_i} )_{i=1}^k }{\type{\rho_c}}{\type{\rho_t}}{\type{\rho_f}}\\ \\
        \norm{f(e_1,...,e_n)} &=& \textrm{let}~
        ([e'_1,...,e'_m]^{\type{\vv{\alpha'}}}, eqs^{\type{\rho_1}}) \leftarrow \norm{e_1,...,e_n} 
         ~~\hfill \textit{Xncall}\\
        && \textrm{in}~ ([x_1{}^{\type{\delta_1}},...,x_k{}^{\type{\delta_k}}],\\
        && ~~(\union{\{ (x_1,...,x_k)= f(e'_1,...,e'_m) \}}{eqs})^{\type{\rho_2}}) \\
 && \textrm{where}~ \type{\rho_2} = \type{\rho [\vv{\alpha'}/\vv{\alpha}] [\vv{\delta}/\vv{\beta}][\gamma'/\gamma]
 } \cup \type{\rho_1} \\
 && \textrm{given}~  \securitySignature{f}{(\type{\vv{\alpha}})}{\type{\vv{\beta}}}
 {\type{\gamma}}{\type{\rho}} ~\textrm{and}~ \type{\gamma'} = \Gamma(\texttt{base}) \\ \\
 \norm{[e_1, \ldots, e_n]} &=& \textrm{let~for}~ i \in \{1,\ldots, n\}:  ~~\hfill \textit{Xtup} \\ 
 && ~~~~~~~~~
 ( [e'_{i 1}{}^{\type{\alpha_{i 1}}}, \ldots, 
   e'_{i m_i}{}^{\type{\alpha_{i m_i}}}], eqs_i^{\type{\rho_i}} ) \leftarrow
 \norm{ e_i }  \\
         && \textrm{in}~ 
          ( [ e'_{11}{}^{\type{\alpha_{1 1}}},
          \ldots, 
          e'_{1 m_1}{}^{\type{\alpha_{1 m_1}}},
          \ldots,
         e'_{k 1}{}^{\type{\alpha_{k 1}}} 
          \ldots, 
          e'_{k m_k}{}^{\type{\alpha_{k m_k}}} 
          ], 
          \\ &&
          ~~~~~
         (\bigcup_{i=1..k} eqs_i)^{\cup_i {\type{\rho_i}}}) \\
\\
\norm{\ttt{base}} &=& \ttt{base}  ~~\hfill \textit{Xbase}\\ 
\norm{\ck{ck}{x=k}} &=& \ck{\norm{ck}}{x=k} ~~\hfill \textit{Xon}\\
\\
 \norm{\vv{x}^{\vv{\beta}} =_{ck^{\type{\gamma}}} \vv{e}} &=& \textrm{let}~
(\vv{e'}{}^{\type{\vv{\alpha}}}, eqs^{\type{\rho}}) \leftarrow \norm{\vv{e}} 
 ~~\hfill \textit{Xeqs}\\
        && \textrm{in}~ 
    (\{ (\vv{x}_j =_{ck} e'_j)_{j=1}^m \} \cup eqs)^{\{ \type{(\gamma \stlub \alpha_i \strel \beta_i)_{i=1}^k} \} \cup \type{\rho}  }
\end{array}
\]
    \caption{\lustre\ to \nlustre\ normalisation}
    \label{FIG:Normalisation}
\end{figure*}

We now present Bourke \textit{et al.}'s ``normalisation transformations'' , which de-nest and distribute operators over lists (tuples) of expressions, and finally transform \ttt{fby} expressions to a form where the first argument is a constant.

Normalising an $n$-tuple of \lustre\ expressions yields an $m$-tuple of \lustre\ expressions without tupling and nesting, and a set of equations, defining fresh local variables (Figure \ref{FIG:Normalisation}).
We denote the transformation as 
\[ ([e'_1, \ldots , e'_m]^{\type{\alpha_1, \ldots, \alpha_m}}, eqs^{\type{\rho}}) \leftarrow \lfloor e_1,..., e_n \rfloor \]
where we have additionally decorated the transformations of \cite{Bourke-jfla2021} with security types for each member of the tuple of expressions, and with a set of type constraints for the generated equations.
We show that the normalisation transformations are indeed \textit{typed transformations}.
Our type annotations indicate why security types and constraints of well-security-typed \lustre\ programs are preserved (modulo satisfaction), as in Theorem \ref{THM:SecPreserve}.

The rules \textit{(Xcnst)}-\textit{(Xbinop)} for constants, variables,  unary and binary operators are obvious, generating no new equations.
In rule \textit{(Xwhn)}, where the sampling condition is distributed over the members of the tuple, the security type for each expression is obtained by taking a join of the security type $\type{\alpha_i}$ of the expression $e'_i$ with $\type{\gamma}$, \textit{i.e.}, that of the variable $x$.

Of primary interest are the rules \textit{(Xfby)} for \texttt{fby} and \textit{(Xncall)} for \textit{node call}, where fresh variables $x_i$ and their defining equations are introduced.
In these cases, we introduce \textit{fresh} security type variables $\type{\delta_i}$ for the $x_i$, and add appropriate constraints.
The rules \textit{(Xite)} and \textit{(Xmrg)} resemble \textit{(Xfby)} in most respects.
In rule \textit{(Xncall)}, the constraints are obtained from the node signature via substitution. 

The rules \textit{(Xbase)} and \textit{(Xon)} for clocks also introduce no equations. 
The rules \textit{(Xtup)} for tuples (lists) of expressions and \textit{(Xeqs)} for equations regroup the resulting expressions appropriately.
The translation of node definitions involves translating the equations, and adding the fresh local variables.

\begin{figure*}
        \begin{equation*}
            \norm{x^{\type{\theta}} =_{ck^{\type{\gamma}}} \lfby{e_0^{\type{\alpha}}}{e^{\type{\beta}}}}_{\textit{fby}} =
              \begin{cases}
                 xinit^{\type{\delta_1}} =_{ck^{\type{\gamma}}} \nlfby{\ttt{true}^{\type{\bot}}}{\ttt{false}^{\type{\bot}}} ~\hfill~ 
                 \type{\bot \stlub \gamma ~\strel~ \delta_1} \\
                 px^{\type{\delta_2}} =_{ck^{\type{\gamma}}} \nlfby{c^{\type{\bot}}}{e^{\type{\beta}}} ~\hfill~ 
                 \type{\gamma \stlub \beta ~\strel~ \delta_2}\\
                 x^{\type{\theta}}  =_{ck^{\type{\gamma}}} \ttt{if}~{xinit^{\type{\delta_1}}}\ttt{then}~{e_0^{\type{\alpha}}} ~\hfill~ 
                 \type{\gamma \stlub \delta_1 \stlub \alpha \stlub \delta_2 ~\strel~ \theta}\\
                 ~~~~~~~~~~~\ttt{else} ~{px^{\type{\delta_2}}}
              \end{cases}       
          \end{equation*}

    \caption{Explicit \ttt{fby} initialisation}
    \label{FIG:L2NLFbyInit}
\end{figure*}

\begin{theorem}[Preservation of security types]
    \label{THM:SecPreserve}
Let $g \in G$ be a node in \lustre\ program $G$.
If the node signature for $g$ in $G$ is 
$\securitySignature{f}{(\vv{\type{\alpha}})}{\vv{\type{\beta}}}{\type{\gamma}}{\type{\rho}}$, correspondingly in $\norm{G}$ it is $ \securitySignature{f}{(\vv{\type{\alpha}})}{\vv{\type{\beta}}}{\type{\gamma}}{\type{\rho'}}$,
and for any ground instantiation $s$,
$s(\type{\rho})$ implies $s(\type{\rho'})$.
\end{theorem}

The proof is on the DAG structure of $G$. 
Here we rely on the modularity of nodes, and the correctness of \type{\textsf{simplify}} (Lemma \ref{LEM:Simplify}).
The proof employs induction on the structure of expressions. 
For the further explicit initialisation of \texttt{fby} (Figure \ref{FIG:L2NLFbyInit}), the preservation of security via \type{\textsf{simplify}} is easy to see.

\paragraph{Semantics preservation.}

We recall the important results from \cite{Bourke-jfla2021}, which establish the preservation of stream semantics by the transformations.

\begin{theorem}[Preservation of semantics. Theorem 2 of \cite{Bourke-jfla2021}]
    \label{THM:JFLA1}
    De-nesting and distribution preserve the semantics of \lustre\ programs. 
    (La passe de d\'{e}simbrication et distributivit\'{e} pr\'{e}serve la s\'{e}mantique des programmes.)
    \[
    \forall G~ f ~\tupstrm{xs}~\tupstrm{ys}:~~~ {\mstrCallPred{G}{\liftnode{f}}{\tupstrm{xs}}{\tupstrm{ys}}} \implies {\mstrCallPred{\norm{G}}{\liftnode{f}}{\tupstrm{xs}}{\tupstrm{ys}}}\]
\end{theorem}

\begin{theorem}[Preservation of semantics. Theorem 3 of \cite{Bourke-jfla2021}]
    \label{THM:JFLA2}
    The explicit initialisations of $\ttt{fby}$ preserve the semantics of the programs. 
    (L’explicitation des initialisations pr\'{e}serve la s\'{e}mantique des programmes.)
    \[
    \forall G~ f ~\tupstrm{xs}~\tupstrm{ys}:~~~ {\mstrCallPred{G}{\liftnode{f}}{\tupstrm{xs}}{\tupstrm{ys}}} \implies {\mstrCallPred{\norm{G}_{fby}}{\liftnode{f}}{\tupstrm{xs}}{\tupstrm{ys}}}\]
\end{theorem}

\subsection{Example}

We adapt an example from \cite{Bourke-jfla2021} to illustrate the translation and security-type preservation.
The \ttt{re\_trig} node in Figure \ref{FIG:L2NLEx} uses the \ttt{cnt\_dn} node (Figure \ref{FIG:L2NLEx1}) to implement a count-down timer that is explicitly triggered whenever there is a rising edge (represented by \ttt{edge})  on \ttt{i}. 
If the count \ttt{v} expires to $0$ before a \ttt{T} on \ttt{i}, the counter isn't allowed restart the count. 
Output \ttt{o} represents an active count in progress. 

We annotate the program with \textcolor{blue}{security types} (superscripts)
and \textcolor{blue}{constraints} for each equation (as comments), according to the typing rules.
\ttt{cnt\_dn} is assumed to have security signature
\(
  \securitySignature{\ttt{cnt\_dn}}
  {(\type{\alpha_1,\alpha_2})}
  {\type{\beta}}
  {\type{\gamma}}
  {\{ \type{\gamma \stlub \alpha_1 \stlub \alpha_2 ~\strel~ \beta} \}}.
\)

Eliminating the security types $\type{\delta'_1}, \type{\delta'_2}, \type{\delta'_3}$,  and $\type{\delta'_6}$, 
of the local variables \ttt{edge}, \ttt{c}, \ttt{v} and nested call to \ttt{cnt\_dn} respectively, we
get the constraint
$\{ \type{\gamma' \lub \alpha'_1 \lub \alpha'_2 \strel \beta'}\}$.

Normalisation introduces local variables (\ttt{v21,v22,v24}) with security types
$\type{\delta'_4}, \type{\delta'_5},  \type{\delta'_6}$.
(Identical names have been used to show the correspondence.)
The $\type{\delta'_i}$ are eliminated by \ckFont{simplify}, and the refinement type $\type{\delta'_6 \{\!| \rho' |\!\}}$ for the node call in the \lustre\ version becomes an explicit constraint $\type{\rho_5}$ in \nlustre.
Observe that the security signature remains the same across the translation.
\begin{figure*}
\footnotesize
    \begin{minipage}{.5\textwidth}
        \begin{lstlisting}[language=Lustre]
node re_trig(i$^{\type{\alpha'_1}}$:bool; n$^{\type{\alpha'_2}}$:int)
  returns (o$^{\type{\beta'}}$ : bool)
  var edge$^{\type{\delta'_1}}$, c$^{\type{\delta'_2}}$:bool,
   v$^{\type{\delta'_3}}$:int;
let
  (edge$^{ck}$)$^{\type{\delta^{'\gamma'}_1}}$ = i$^{\type{\alpha'_1}}$ and 
    (false$^{\type{\bot}}$ fby (not i$^{\type{\alpha'_1}}$)); 
-- $\type{\rho_{1L}}= \{\type{\bot \lub \alpha'_1 \lub \bot \lub \alpha'_1 \rel \delta'_1}\}$
  (c$^{ck}$)$^{\type{\delta^{'\gamma'}_2}}$ = edge$^{\type{\delta'_1}}$ or 
    (false$^{\type{\bot}}$ fby o$^{\type{\beta'}}$);
-- $\type{\rho_{2L}}= \{\type{\gamma' \lub \delta'_1 \lub \bot \lub \beta' \rel \delta'_2}\}$
  (v$^{c}$)$^{\type{\delta^{'\delta'_2}_3}}$ = merge c$^{\type{\delta'_2}}$
   (cnt_dn((edge$^{\type{\delta'_1}}$, n$^{\type{\alpha'_2}}$)
    when c$^{\type{\delta'_2}}$))$^{{\type{\delta'_6\{|\rho'|\}}}^{\type{\delta'_2}}}$ 
   (0 when not c$^{\type{\delta'_2}}$);
-- $\type{\rho'}= \{ \type{\delta'_2 \lub (\delta'_1 \lub \delta'_2) \lub (\alpha'_2 \lub \delta'_2) \rel\delta'_6}\}$
-- $\type{\rho_{3L}}= \{\type{\delta'_2 \lub \delta'_2 \lub \delta'_6 \lub \bot \lub \delta'_2 \rel \delta'_3}\} \cup \type{\rho'}$
   (o$^{c}$)$^{\type{\beta^{'\delta'_2}}}$= v$^{\type{\delta'_3}}$ > 0$^{\type{\bot}}$;
-- $\type{\rho_{4L}}= \{\type{\delta'_2 \lub \delta'_3 \lub \bot \rel \beta'}\}$
tel
    \end{lstlisting}
    \end{minipage}%
        \begin{minipage}{.5\textwidth}

\begin{lstlisting}[language=Lustre]
node re_trig(i$^{\type{\alpha'_1}}$:bool; n$^{\type{\alpha'_2}}$:int)
 returns (o$^{\type{\beta'}}$ : bool)
 var edge$^{\type{\delta'_1}}$, ck$^{\type{\delta'_2}}$:bool, v$^{\type{\delta'_3}}$:int,
  v22$^{\type{\delta'_4}}$:bool, v21$^{\type{\delta'_5}}$:bool,
  v24$^{\type{\delta'_6}}$:int when ck;
let
  v22$^{{\type{\delta_4}}}$  =$_{\type{\delta'_2}}$ false$^{\type{\bot}}$ fby 
    (not i$^{\type{\alpha'_1}}$);
-- $\type{\rho_1} = \{\type{\delta'_2 \lub \bot \lub \alpha'_1 \rel \delta'_4}\} $ 
  edge$^{\type{\delta'_1}}$ =$_{\type{\bot}}$ i$^{\type{\alpha'_1}}$ and v22$^{\type{\delta'_4}}$;
-- $\type{\rho_2} = \{\type{ \bot \lub \alpha'_1 \lub \delta'_4 \rel \delta'_1}\} $ 
  v21$^{\type{\delta'_5}}$ =$_{\type{\bot}}$ false$^{\type{\bot}}$ fby o$^{\type{\beta'}}$;
-- $\type{\rho_3} = \{\type{\bot \lub \bot \lub \beta' \rel \delta'_5}\}$
  ck$^{\type{\delta'_2}}$  =$_{\type{\gamma'}}$ edge$^{\type{\delta'_1}}$ or v21$^{\type{\delta'_5}}$;
-- $\type{\rho_4} = \{\type{\bot \lub \delta'_1 \lub \delta'_5 \rel \delta'_2}\}$
  v24$^{\type{\delta'_6}}$ =$_{\type{\delta'_2}}$ cnt_dn(
    edge$^{\type{\delta'_1}}$ when ck$^{\type{\delta'_2}}$,
   n$^{\type{\alpha'_2}}$ when ck$^{\type{\delta'_2}}$);
-- $\type{\rho_5} = \{\type{\delta'_2 \lub (\delta'_1 \lub \delta'_2) \lub (\alpha'_2 \lub \delta'_2) \rel \delta'_6}\}$
  v$^{\type{\delta'_3}}$ =$_{\type{\delta'_2}}$ merge ck$^{\type{\delta'_2}}$ v24$^{\type{\delta'_6}}$ 
    (0$^{\type{\bot}}$ when not ck$^{\type{\delta'_2}}$);
-- $\type{\rho_6}= \{\type{\delta'_2 \lub \delta'_2 \lub \delta'_6 \lub \bot \lub \delta'_2 \rel \delta'_3}\}$
  o$^{\type{\beta'}}$ =$^{\type{\delta'_2}}$ v$^{\type{\delta'_3}}$>0$^{\type{\bot}}$;
-- $\type{\rho_7}= \{\type{\delta'_2 \lub \delta'_3 \lub \bot \rel \beta'}\}$
tel
\end{lstlisting}
        \end{minipage}
        \noindent\rule{\linewidth}{0.4pt}
        \centering
        \begin{align*}
        \lsimpl{(\type{\beta'}, \{ \union{\type{\rho_{1L}}}{\type{\rho_{2L}}}{\type{\rho_{3L}}}{\type{\rho_{4L}}}\})}{\{\type{\delta'_1},\type{\delta'_2},\type{\delta'_3}, \type{\delta'_6}\}} = 
        (\type{\beta'},
        \{ \type{\gamma' \lub \alpha'_1 \lub \alpha'_2 \rel \beta'}\}) \\
        \nlsimpl{(\type{\beta'}, \{ \union{\type{\rho_{1}}}{\type{\rho_{2}}}{\type{\rho_{3}}}{\type{\rho_{4}}}{\type{\rho_{5}}}{\type{\rho_{6}}}{\type{\rho_{7}}}\})}{\{\type{\delta'_1},\type{\delta'_2},\type{\delta'_3}, \type{\delta'_4}, \type{\delta'_5}, \type{\delta'_6}\}} \\
        \hfill = (\type{\beta'}, \{ \type{\gamma' \lub \alpha'_1 \lub \alpha'_2 \rel \beta'}\})
      \end{align*}
        \caption{Example: Security analysis and normalisation.  \ttt{when  c} and \ttt{when not c} abbreviate \ttt{when c = T} and \ttt{when c = F}.}
        \label{FIG:L2NLEx}
\end{figure*}

%% file: noninterference.tex
\section{Security and Non-Interference}\label{SEC:noninterference}

We first recall and adapt concepts from our previous work \cite{PYS-memocode2020}.

\begin{lemma}[Security of Node Calls; \textit{cf}. Lemma 3 in \cite{PYS-memocode2020}]\label{LEM:sec-nodecall}
Assume the following, for a call to a node with the given security signature
\[ \securitySignature{f}{(\vv{\type{\alpha}})}{\vv{\type{\beta}}}{\type{\gamma}}{\type{\rho}} ~~~\hfill~~~
\mexprPred{\Gamma}{\vv{e}}{\vv{\type{\alpha'}}}
~~~\hfill~~~
\mexprPred{\Gamma}{ f(\vv{e})}{\vv{\beta'}}
~~~\hfill~~~
\mclkPred{\Gamma}{ck}{\type{\gamma}}
\]
where $ck$ is the base clock underlying the argument streams $\vv{e}$.
Let $s$ be a ground instantiation of type variables such that for some security classes $\vv{t},w \in \tti{SC}$:
$s(\type{\vv{\alpha'}}) = \vv{t}$ 
and $s(\type{\gamma})=w$. \\[1mm]
Now, if $\type{\rho}$ is satisfied by the ground instantiation $\{ 
\type{\vv{\alpha}} \mapsto  \vv{t},
\type{\vv{\beta}} \mapsto  \vv{u},
\type{\gamma} \mapsto w \}$, 
then the  $s(\type{\vv{\beta'}})$ are defined, and 
$s(\type{\vv{\beta'}}) ~\rel~ s(\type{\vv{\beta} \{\!| \type{\rho} |\!\})}$.
\end{lemma}
Lemma \ref{LEM:sec-nodecall} relates the satisfaction of constraints on security types generated during a node call to satisfaction in  a security lattice via a ground instantiation.
Again we rely on the modularity of nodes --- that no recursive calls are permitted, and nodes do not have free variables.

\begin{definition}[Node Security;
Definition III.1 in \cite{PYS-memocode2020}]
\label{DEF:nodesec}
Let $g$ be a node in the program graph $G$ with security signature
\( \securitySignature{f}{(\vv{\type{\alpha}})}{\vv{\type{\beta}}}{\gamma}{\type{\rho}}. 
\)
Let $s$ be a ground instantiation that maps the security type variables in the set
$\set{\type{\vars{\alpha}{n}}} \cup \set{\type{\vars{\beta}{m}}} \cup \set{\type{\gamma}}$
to the security class lattice $\tti{SC}$. \\
Node $g$ is secure with respect to $s$ if 
   (i)  $\type{\rho}$ is satisfied by $s$;
    (ii) For each node $g'$ on which $g$ is directly dependent, $g'$ is secure with respect to the appropriate ground instantiations for each call to $g'$ in $g$ as given by Lemma \ref{LEM:sec-nodecall}.
\end{definition}
This definition captures the intuition of node security in that all the constraints generated for the equations within the node must be satisfied, and that each internal node call should also be secure.  

\ignore{
By induction on the structure of expressions, it is easy to prove that the security level of expressions is at least as high as those of its constituent variables.
\begin{lemma}[Simple Security]
    \label{LEM:SimpleSec}
    For any general expression $e$ and security type assumption $\Gamma$, if 
    $\mexprPred{\Gamma}{e}{\alpha}$, then for all $x \in fv(e): s(\Gamma(x)) \rel s(\type{\alpha})$, for all $s$.
    \end{lemma}
}   
The notion of non-interference requires limiting observation to streams whose security level is at most a given security level $t$.

\begin{definition}[$(\rel t)$-projected Stream; Definition IV.1 in \cite{PYS-memocode2020}]
    \label{DEF:alphaProj}
    Suppose $t \in \tti{SC}$ is a security class.
    Let $X$ be a set of program variables, $\Gamma$ be security type assumptions for variables in $X$, and $s$ be a ground instantiation,
    i.e., $\Gamma \circ s$ maps variables in $X$ to security classes in 
    $\tti{SC}$.
    Let us define $X_{\rel t} = \{ x \in X ~|~ (\Gamma \circ s)(x) \rel t \}$.
    Let $\Hst$ be a Stream history such that $X \subseteq \tti{dom}(\Hst)$.
    Define $\Hst |_{X_{\rel t}}$ as the projection of $\Hst$ to $X_{\rel t}$, \textit{i.e.}, restricted to those variables that are at security level $t$ or lower:
    \[
    \Hst |_{X_{\rel t}}  (x) = \Hst (x) ~~~~\mbox{for $x \in X_{\rel t}$}.
    \]
\end{definition}

\begin{theorem}[Non-interference for \nlustre; Theorem 5 in \cite{PYS-memocode2020}]
\label{THM:Non-interference}
    Let $g \in G$ be a node with security signature
        \[ \securitySignature{f}{\vv{\type{\alpha}}}{\vv{\type{\beta}}}{\type{\gamma}}{\type{\rho}} \]
    which is secure with respect to ground instantiation $s$ of the type variables. \\
    Let $eqs$ be the set of equations in $g$. 
    Let $X = fv(eqs)-dv(eqs)$, \textit{i.e.}, the input variables in $eqs$. \\
    Let $V = fv(eqs) \cup dv(eqs)$, \textit{i.e.}, the input, output and local variables. \\
    Let $\Gamma$ (and $s$) be such that 
    \(
    \meqnPred{\Gamma}{ eqs }{\rho}
    \)
    and $\type{\rho}$ is satisfied by $s$.
    Let $t \in \tti{SC}$ be any security level. 
    Let $bs$ be a given (base) clock stream. \\
    Let $\Hst$ and $\Hst'$ be such that
    \begin{enumerate}
        \item for all $eq \in eqs$:  $\mstrEqnPred{G}{\Hst}{bs}{eq}$
        and $\mstrEqnPred{G}{\Hst'}{bs}{eq}$, \textit{i.e.}, both $\Hst$ and $\Hst'$ are consistent Stream histories on each of the equations.
        \item $\Hst |_{X_{\rel t}} = \Hst' |_{X_{\rel t}}$, i.e., 
        $\Hst$ and $\Hst'$ agree on the input variables which are at a security level $t$ or below.
    \end{enumerate}
    Then $\Hst |_{V_{\rel t}} = \Hst' |_{V_{\rel t}}$, i.e, $\Hst$ and $\Hst'$ agree on all variables of the node that are given a security level $t$ or below.
\end{theorem}

\begin{theorem}[Non-interference for \lustre]\label{THM:soundness}
If program $G$ is well-security-typed in \lustre, then it exhibits non-interference with respect to \lustre's stream semantics.
\end{theorem}
\label{THM:Derived-Non-interference}
\Proofsketch \ 
Let $G$ be well-security-typed in \lustre.
This means that each node $g \in G$ is well-security-typed.
By induction on the DAG structure of $G$, using Theorem \ref{THM:SecPreserve}, $\norm{G}$ is well-security-typed.
By Theorem \ref{THM:Non-interference}, $\norm{G}$ exhibits non-interference.
By Theorems \ref{THM:JFLA1} and \ref{THM:JFLA2}, $\norm{G}$ and $G$ have the same \textit{extensional} semantics for each node.
Therefore, $G$ exhibits non-interference.

%% file: conclusions.tex
\section{Conclusions}
\label{SEC:Conclusion}

We have presented a novel security type system for \lustre\ using the notion of constraint-based refinement (sub)types.  
Using security-type preservation and earlier results, we have shown its semantic soundness, expressed in terms of non-interference, with respect to the language's stream semantics.

We are developing mechanised proofs of these results, which can be integrated into the Vel\'{u}s verified compiler framework \cite{Velus}.

While \lustre's value type system is quite jejune, this security type system is not.
It is therefore satisfying to see that it satisfies a subject reduction property\footnote{
At SYNCHRON 2020, De Simone asked Jeanmaire and Pesin whether the terminology ``normalisation'' used in their work  \cite{Bourke-jfla2021} was related in any way to notions of normalisation seen in, \textit{e.g.}, the $\lambda$-calculus. It is!}.
A difficult aspect encountered during the transcription phase  \cite{Bourke-jfla2021} concerns alignment of clocks in the presence of complex clock dependencies. 
We clarify that our type system, being static, only considers \type{security levels of clocks}, not actual clock behaviour, and therefore is free from such complications.
Further, the clocks induce no timing side-channels since the typing rules enforce, \textit{a fortiori}, that the security type of any (clocked) expression is at least as high as that of its clock.

\ignore{

}

%% file: APP_free_vars.tex
\section{Free Variable Definitions}
\label{APP:FVDef}
The definitions of free variables ($fv$) in expressions and equations, and defined variables ($dv$) in equations are given in \autoref{FIG:FVEdef} and \autoref{FIG:FVEQdef}.

\begin{figure}[ht]
\[
\begin{array}{rcl}
    fv(c) &=& \{\}  \\
    fv(x) &=& \{x\}  \\
    fv(\unop{e})&=& fv(e) \\
    fv(\binop{e_1}{e_2})&=& \union{fv(e_1)}{fv(e_2)}\\
    fv(\when{\vv{e}}{x=k})&=&\union{fv(\vv{e})}{\set{x}}\\
    fv(\lmerge{x}{\vv{e_1}}{\vv{e_2}}) &=& 
     \union{\set{x}}{fv(\vv{e_1})}{fv(\vv{e_2})} \\
    fv(\fby{\vv{e_1}}{\vv{e_2}}) &=& 
     \union{fv(\vv{e_1})}{fv(\vv{e_2})} \\
    fv(\ite{e_1}{\vv{e_2}}{\vv{e_3}}) &=& 
    \union{fv(e_1)}{fv(\vv{e_2})}{fv(\vv{e_3})} \\
    fv(f(\vv{e})) &=& \bigcup_{i} fv(\vv{e_i}) \\ \\[1mm]
    fv(\ttt{base}) &=& \{\ttt{base}\} \\
    fv(\ck{ck}{x=k}) &=& \union{fv(ck)}{\set{x}} \\ \\[1mm]
    fv(\vv{e})  &=& \bigcup_i {fv(e_i)} \\ \\[1mm]
    fv(e::ck) &=& \union{fv(e)}{fv(ck)} 
\end{array}
\]
    \caption{Free variables for expressions}
    \label{FIG:FVEdef}
\end{figure}

\begin{figure}[ht]
\[
\begin{array}{rcl}
    fv(\vv{x}=\vv{e} ) &=& fv(\vv{e}) \setminus \{\vv{x} \} \\
    \\[1mm]
    dv(\vv{x} =\vv{e}) &=& \{\vv{x}\}\\
\end{array}
\]

\[
\begin{array}{rcl}
    fv(\eqn{x}{ce}{ck}) &=& \union{fv(ck)}{fv(ce)} \setminus \{x\} \\
    fv(\eqn{x}{\fby{c}{e}}{ck}) &=& \union{fv(ck)}{fv(e)} \setminus \{x\} \\
    fv(\eqn{\vv{x}}{f(\vv{e})}{ck}) &=& \union{fv(ck)}{fv(\vv{e})} \setminus \set{\vv{x}} \\ 
    \\[1mm]
 dv(\eqn{x}{ce}{ck}) &=&  \{x\} \\
    dv(\eqn{x}{\fby{c}{e}}{ck}) &=&  \{x\} \\
    dv(\eqn{\vv{x}}{f(\vv{e})}{ck}) &=& \set{\vv{x}} 
\end{array}
\]
    \caption{Free and defined variables for equations}
    \label{FIG:FVEQdef}
\end{figure}

\ignore{
\begin{figure*}[ht]
\begin{minipage}{.3\textwidth}
\begin{align*}
    &fv(x)~ = ~ \{x\}  \\
    &fv(c)~ =~ \{\}  \\
    &fv(\unop{e})~ =~ fv(e) \\
    &fv(\binop{e_1}{e_2})~=~ \union{fv(e_1)}{fv(e_2)}\\
    &fv(\when{e}{x=k})~=~\union{fv(e)}{\set{x}}\\
    &fv(\ttt{base})~=~ \{\ttt{base}\} \\
\end{align*}

\end{minipage}%
\begin{minipage}{.7\textwidth}
\begin{align*}
    &fv(\vv{e})  ~=~ \bigcup_i {fv(e_i)} \\
    &fv(\lmerge{x}{e_1}{e_2}) ~ =~ \\
    &\qquad \union{\set{x}}{fv(e_1)}{fv(e_2)} \\
    &fv(\ite{e_1}{e_2}{e_3})~ =~ \\
    &\qquad \union{fv(e_1)}{fv(e_2)}{fv(e_3)}\\
    &fv(\ck{ck}{x=k}) ~=~ \union{fv(ck)}{\set{x}} \\
\end{align*}

\end{minipage}
\begin{minipage}{.4\textwidth}
\begin{align*}
&dv(\eqn{x}{ce}{ck}) ~=~  \{x\} \\
    &dv(\eqn{x}{\fby{c}{e}}{ck}) ~=~  \{x\} \\
    &dv(\eqn{x}{f(\vv{e})}{ck}) ~=~ \set{\vv{x}} \\
\end{align*}
\end{minipage}%
\begin{minipage}{.7\textwidth}
\begin{align*}
    &fv(\eqn{x}{ce}{ck}) ~=~ \union{fv(ck)}{fv(ce)} \setminus \{x\} \\
    &fv(\eqn{x}{\fby{c}{e}}{ck}) ~=~ \union{fv(ck)}{fv(e)} \setminus \{x\} \\
    &fv(\eqn{x}{f(\vv{e})}{ck}) ~=~ \union{fv(ck)}{fv(\vv{e})} \setminus \set{\vv{x}} \\
\end{align*}
\end{minipage}%
    \caption{Definition of $fv$ and $dv$ functions}
    \label{FIG:FVdef}
\end{figure*}
}

%% file: APP_consolidated_stream_semantics.tex
\section{Stream Semantics}
\label{APP:LustreStreamSem}

We present here a specification of core \lustre's co-inductive stream semantics, with some commentary and intuition.
This consolidates various earlier presentations of rules \cite{SNLustre2017,Lustre2020,PYS-memocode2020,Bourke-jfla2021}, and can be seen as an abstract Coq-independent specification of the semantics encoded in  the V\'{e}lus development.

\begin{figure*}
    $$
    \begin{array}{c}
        \namedJdg{\semConst{c}{bs}{cs}}
        {\mstrSemExpPred{G,\Hst}{bs}{c}{[cs]}}{LScnst} ~~~
        \namedJdg{\Hst(x)={\textcolor{BrickRed}{xs}}}
        {\mstrSemExpPred{G,\Hst}{bs}{x}{[xs]}}{LSvar} \\
        \\
        \namedJdg{\predSet{\mstrSemExpPred{G,\Hst}{bs}{e}{[es]}}
            {\liftunop{es}{os} }}
        {\mstrSemExpPred{G,\Hst}{bs}{\unop{e}}{[os]}}{LSunop} \\
        \\
        \namedJdg{\predSet{\mstrSemExpPred{G,\Hst}{bs}{e_1}{[es_1]}}
            {\mstrSemExpPred{G,\Hst}{bs}{e_2}{[es_2]}}
            {\liftbinop{es_1}{es_2}{os} }}
        {\mstrSemExpPred{G,\Hst}{bs}{\binop{e_1}{e_2}}{[os]}}{LSbinop} \\
        \\
        \namedJdg{\predSet{\forall i ~\mstrSemExpPred{G,\Hst}{bs}{e_i}{\tupstrm{es}_i}}
            {\hist{x}={\textcolor{BrickRed}{xs}}}
            {\forall i:~\mapwhenk{k}{xs}{\tupstrm{es}_i}{\tupstrm{os}_i}}}
        {\mstrSemExpPred{G,\Hst}{bs}{\when{\vv{e_i}}{x=k}}{
        \flatten{\vv{\tupstrm{os}_i}}}}{LSwhn}\\
        \\
        \namedJdg{\dependSet{\predSet{\mstrSemExpPred{G,\Hst}{bs}{e}{[es]}}
        {\forall i:~ ~\mstrSemExpPred{G,\Hst}{bs}{et_i}{\tupstrm{ets}_i}}
        }
        {
           \forall j:~  \mstrSemExpPred{G,\Hst}{bs}{ef_j}{\tupstrm{efs}_j} 
            ~~~~
            \mapsemIte{es}{(\flatten{\vv{\tupstrm{ets}_i}})}{
            (\flatten{\vv{\tupstrm{efs}_j}})}{\tupstrm{os}}
            }}
        {\mstrSemExpPred{G,\Hst}{bs}{\ite{e}{\vv{et_i}}{\vv{ef_j}}}{\tupstrm{os}}}{LSite} \\
        \\
       \namedJdg{\dependSet{\predSet{\hist{x}=xs}{
       \forall i: ~\mstrSemExpPred{G,\Hst}{bs}{et_i}{\tupstrm{ets}_i}
       }}
            {\forall j: ~
            \mstrSemExpPred{G,\Hst}{bs}{ef_j}{\tupstrm{efs}_j}
            ~~~~
             \mapsemMerge{xs}{(\flatten{\vv{\tupstrm{ets}_i}})}{
            (\flatten{\vv{\tupstrm{efs}_j}})}{\tupstrm{os}}}
            }
       {\mstrSemExpPred{G,\Hst}{bs}{\lmerge{x}{\vv{et_i}}{\vv{ef_j}}}{\tupstrm{os}}}{LSmrg} \\
       \\
       \namedJdg{\dependSet{\predSet{\forall i:~ ~\mstrSemExpPred{G,\Hst}{bs}{e0_i}{\tupstrm{e0s}_i}
      ~~~~ \forall j:~ 
       \mstrSemExpPred{G,\Hst}{bs}{e_j}{\tupstrm{es}_j}}}
            { ~\mapsemLFby{(\flatten{\vv{\tupstrm{e0s}_i}})}{
            (\flatten{\vv{\tupstrm{es}_j}})}{\tupstrm{os}}}}
        {\mstrSemExpPred{G,\Hst}{bs}{\fby{\vv{e0_i}}{\vv{e_j}}}{\tupstrm{os}}}{LSfby} \\
   \\
        \namedJdg{\forall i \in [1,k] ~ \mstrSemExpPred{G,\Hst}{bs}{e_i}{\tupstrm{es}_i} ~~~
        [\hist{x_1}, \ldots, \hist{x_n}] = 
        \flatten{\vv{\tupstrm{es}_i}}
        }
        {\mstrEqnPred{G}{\Hst}{bs}{\vv{x_j}=\vv{e_i}}}{LSeq} \\
  \\
    \namedJdg{\dependSet{\predSet{\node \in G}
        {\hist{n.\tbf{in}} = \tupstrm{xs}}
        {\baseOf{\tupstrm{xs}} = bs}}
        {\predSet{\hist{n.\tbf{out}} = \tupstrm{ys}}
        {\forall eq \in \vv{eq}:~ \mstrEqnPred{G}{\Hst}{bs}{eq}}
         }
    }{\mstrCallPred{G}{\liftnode{f}}{\tupstrm{xs}}{\tupstrm{ys}}}{LSndef} \\
    \\
           \namedJdg{\predSet{\mstrSemExpPred{G,\Hst}{bs}{\vv{e_i}}{\tupstrm{xs}}}
            {\mstrCallPred{G}{\liftnode{f}}{\tupstrm{xs}}{\tupstrm{ys}}}}
        {\mstrSemExpPred{G,\Hst}{bs}{f(\vv{e_i})}{\tupstrm{ys}}}{LSncall}
    \end{array}
    $$
    \caption{Stream semantics of \lustre}
    \label{FIG:APPLusStrSemNodeEqn}
\end{figure*}

The semantics of \lustre\ and \nlustre\  programs are \textit{synchronous}:
Each variable and expression defines a data stream which pulses with respect to a \textit{clock}.
A clock is a stream of booleans (CompCert/Coq's $\true$ and $\false$ in Velus).
A flow takes its $n^\tti{th}$ value on the $n^\tti{th}$ clock tick, \textit{i.e.},  some value, written $\stream{v}$, is present at instants when the clock value is \true, and none (written $\nullStream)$ when it is \false.
The \textit{temporal operators} \ttt{when}, \ttt{merge} and \ttt{fby} are used to express the complex clock-changing and clock-dependent behaviours of sampling, interpolation and delay respectively.


Formally the stream semantics is defined using predicates over the program graph $G$, a (co-inductive) stream \textit{history} ($\Hst: \tti{Ident} \rightarrow \tti{value}~\tti{Stream}$) that associates value streams to variables, and a clock $bs$ \cite{Lustre2020,PYS-memocode2020,Bourke-jfla2021}.
Semantic operations on (lists of) streams are written in \textcolor{blue}{blue \textsf{sans serif}} typeface.
Streams are written in \textcolor{BrickRed}{red}, with lists of streams usually written in \textbf{\textcolor{BrickRed}{bold face}}.
All these stream operators, defined co-inductively,  enforce the clocking regime, ensuring  the presence of a value when the clock is \ckFont{true}, and absence when \ckFont{false}.

The predicate $\mstrSemExpPred{G,\Hst}{bs}{e}{\tupstrm{es}}$ relates an \textit{expression} $e$ to a \textit{list} of streams, written $\tupstrm{es}$.
A list consisting of only a single stream $\textcolor{BrickRed}{es}$ is explicitly denoted as $\textcolor{BrickRed}{[es]}$.
The semantics of \textit{equations} are expressed using the predicate 
$\mstrEqnPred{G}{\Hst}{bs}{\vv{eq_i}}$, which requires \textit{consistency} between the assumed and defined stream histories in $\Hst$ for the program variables, as induced by the equations.
Finally, the semantics of \textit{nodes} is given as a stream history transformer predicate
$\mstrCallPred{G}{\liftnode{f}}{\tupstrm{xs}}{\tupstrm{ys}}.$

Figure \ref{FIG:APPLusStrSemNodeEqn} presents the stream semantics for \lustre.
While rules for \textit{some} constructs have been variously presented \cite{SNLustre2017,Lustre2020,PYS-memocode2020,Bourke-jfla2021}, our presentation can be considered as a definitive consolidated specification of the operational semantics of \lustre, consistent with the V\'{e}lus compiler encoding \cite{Velus}.

Rule (LScnst) states that a constant $c$ denotes a constant stream of the value $\stream{c}$ pulsed according to given clock $bs$.  
This is effected by the semantic operator \ckFont{const}.
Rule (LSvar) associates a variable $x$ to the stream given by $\Hst(x)$. 
In rule (LSunop), \ckFont{$\hat{\diamond}$} denotes the operation $\diamond$ \textit{lifted} to apply instant-wise to the stream denoted by expression $e$.
Likewise in rule (LSbinop), the  binary operation $\oplus$ is applied paired point-wise to the streams denoted by the two sub-expressions (which should both pulse according to the same clock).
In all these rules, an expression is associated with a \textit{single} stream. 

The rule (LSwhn) describes \textit{sampling} whenever a variable $x$ takes the boolean value $k$, from the flows arising from a list of expressions $\vv{e_i}$,  returning a list of streams of such sampled values.
The predicate $\widehat{\ckFont{when}}$ \textit{maps} the predicate $\ckFont{when}$ to act on the corresponding components of \textit{lists} of streams, \textit{i.e.},  \[ 
\mapwhenk{k}{xs}{[es_1, \ldots, es_k]}{[os_1, \ldots, os_k]} 
~\textrm{abbreviates} ~ 
\bigwedge_{i \in [1,k]}~ \whenk{k}{xs}{es_i}{os_i}.
\]
(Similarly for the predicates
$\widehat{\ckFont{merge}}$, $\widehat{\ckFont{ite}}$, and $\widehat{\ckFont{fby}_L}$.  
The operation $\flatten{\_}$ flattens a list of lists (of possibly different lengths) into a single list. 
Flattening is required since expression $e_i$ may in general denote a \textit{list} of streams \textcolor{BrickRed}{$\tupstrm{es}_i$}.

The expression $\lmerge{x}{\vv{et}_i}{\vv{ef}_j}$ achieves (lists of) streams on a faster clock.
The semantics in rule (LSmrg) assume that for each pair of corresponding component streams from
$\flatten{\tupstrm{ets}_i}$ and
$\flatten{\tupstrm{efs}_j}$, a value is present in the first stream and absent in the second at those instances where $x$ has a true value $\stream{T}$, and complementarily, a value is present in the second stream and absent in the first when $x$ has a false value $\stream{F}$.
Both values must be absent when $x$’s value is absent.
These conditions are enforced by the auxiliary semantic operation \ckFont{merge}. 
In contrast, the conditional expression $\ite{e}{\vv{et}}{\vv{ef}}$
requires that all three argument streams $es$, and the corresponding components from $\flatten{\vv{\tupstrm{ets}_i}}$ and
$\flatten{\vv{\tupstrm{efs}_j}}$ pulse to the same clock.
Again, values are selected from the first or second component streams depending on whether the stream $es$ has the value $\stream{T}$ or $\stream{F}$ at a particular instant.
These conditions are enforced by the auxiliary semantic operation \ckFont{ite}. 
A delay operation is implemented by $\fby{e0}{e}$.
The rule (LSfby) is to be read as follows.
Let each expression $e0_i$ denote a list of streams
$\tupstrm{e0s_i}$, and each expression $e_j$ denote a list of streams $\tupstrm{es_j}$.
The output list of streams consists of streams whose first elements are taken from the each stream in $\flatten{\vv{\tupstrm{e0s}_i}}$ with the rest taken from the corresponding component of $\flatten{\vv{\tupstrm{es}_j}}$.
These are achieved using the semantic operation \textcolor{blue}{\textsf{fby$_L$}}.

The rule (LSeq) for equations checks the consistency between the assumed meanings for the defined variables $x_j$ according to the history $\Hst$ with the corresponding components of the list of streams $\flatten{\vv{\tupstrm{es}_i}}$ to which a tuple of right-hand side expressions evaluates.

The rule (LSndef) presents the meaning given to the definition named $f$ of a node as a stream list transformer $\liftnode{f}$. 
If history $\Hst$ assigns lists of streams to the input and output variables for a node in a manner such that the semantics of the equations in the node are satisfied, then the semantic funnction $\liftnode{f}$ transforms input stream list $\tupstrm{xs}$ to output stream list $\tupstrm{ys}$.
The operation \ckFont{base-of} finds an appropriate base clock with respect to which a given list of value streams pulse.
The rule (LSncall) applies the stream transformer semantic function $\liftnode{f}$ to the stream list $\tupstrm{xs}$ corresponding to the tuple of arguments $\vv{e_i}$, and returns the stream list $\tupstrm{ys}$.

\paragraph{Clocks and clock-annotated expressions.} \ 
We next present rules for clocks.
Further, we  assume that all (\nlustre) expressions in equations can be clock-annotated, and present the corresponding rules.

\begin{figure*}
$$
   \begin{array}{c}
        \namedJdg{}
        {\mstrSemCkPred{\Hst}{bs}{\ttt{base}}{bs}}{LSbase}  \\ \\
        \namedJdg{
            \dependSet{\predSets{\mstrSemCkPred{\Hst}{bs}{ck}{\cc{\true}{bk}}}
            {\hist{x}=\cc{\stream{k}}{xs}}}
            {\mstrSemCkPred{\htl{\Hst}}{\tl{bs}}{\on{ck}{x}{k}}{bs'}}}
        {\mstrSemCkPred{\Hst}{bs}{\on{ck}{x}{k}}{\cc{\true}{bs'}}}{LSonT} \\

        \namedJdg{
            \dependSet{\predSets{\mstrSemCkPred{\Hst}{bs}{ck}{\cc{\false}{bk}}}
            {\hist{x}=\cc{\nullStream}{xs}}}
            {\mstrSemCkPred{\htl{\Hst}}{\tl{bs}}{\on{ck}{x}{k}}{bs'}}}
        {\mstrSemCkPred{\Hst}{bs}{\on{ck}{x}{k}}{\cc{\false}{bs'}}}{LSonA1} 
        \\
               \namedJdg{
            \dependSet{
             \predSet{\mstrSemCkPred{\Hst}{bs}{ck}{\cc{\true}{bk}}}
                {\hist{x}=\cc{\stream{k}}{xs}}
            }
            {\mstrSemCkPred{\htl{\Hst}}{\tl{bs}}{\onF{ck}{x}{k}}{bs'}}}
        {\mstrSemCkPred{\Hst}{bs}{\onF{ck}{x}{k}}{\cc{\false}{bs'}}}{LSonA2} \\ \\

        \namedJdg{\dependSet{\mstrSemExpPred{\Hst}{bs}{e}{[\ccnb{\nullStream}{es}]}}
        {\mstrSemCkPred{\Hst}{bs}{ck}{\ccnb{\false}{cs}}}}
    {\mstrSemExpPred{\Hst}{bs}{e::ck}{[\ccnb{\nullStream}{es}]}}{NSaeA} ~~
    
        \namedJdg{\dependSet{\mstrSemExpPred{\Hst}{bs}{e}{[\ccnb{\stream{v}}{es}]}}
            {\mstrSemCkPred{\Hst}{bs}{ck}{\ccnb{\true}{cs}}}}
        {\mstrSemExpPred{\Hst}{bs}{e::ck}{[\ccnb{\stream{v}}{es}]}}{NSae} 
    \end{array}
$$   
    \caption{Stream semantics of clocks and annotated expressions}
    \label{FIG:StrSemCk}
\end{figure*}

The predicate 
$\mstrSemCkPred{\Hst}{bs}{ck}{bs'}$ in \autoref{FIG:StrSemCk} defines the meaning of a \nlustre\ clock expression $ck$ with respect to a given history $\Hst$ and a clock $bs$ to be the resultant clock $bs'$.
The \ttt{on} construct lets us define 
coarser clocks derived from  a given clock --- whenever a variable $x$ has the desired value $k$ and the given clock is true. 
The rules (LSonT), (LSonA1), and (LSonA2) 
present the three cases: respectively when variable $x$ has the desired value $k$ and clock is true; the clock is false; and  the program variable $x$ has the complementary value and the clock is true.
The auxiliary operations \ckFont{tl} and \ckFont{htl},  give the tail of a stream, and the tails of streams for each variable according to a given history $\Hst$. 
Rules (NSaeA)-(NSae) describe the semantics of  clock-annotated expressions, where the output stream carries a value exactly when the clock is true.

\paragraph{Stream semantics for \nlustre.} \ \ 
The semantic relations for \nlustre\ are either identical to (as in constants, variables, unary and binary operations) or else the (simple) singleton cases of the rules given for \lustre\ (as in \ttt{merge}, \ttt{ite}, \ttt{when}).

The significant differences are in treatment of \ttt{fby}, and the occurrence of \ttt{fby} and node call only in the context of equations.

\begin{figure*}
   $$
    \begin{array}{c}
    \namedJdg{\dependSet{\predSet{\node \in G}
        {\hist{n.\tbf{in}} = \tupstrm{xs}}
        {\baseOf{\tupstrm{xs}} = bs}}
        {\predSet{\resClk{\Hst}{bs}}
        {\hist{n.\tbf{out}} = \tupstrm{ys}}
        {\forall eq \in \vv{eq}:~ \mstrEqnPred{G}{\Hst}{bs}{eq}}
         }
    }{\mstrCallPred{G}{\liftnode{f}}{\tupstrm{xs}}{\tupstrm{ys}}}{NSndef'} \\
    \\
        \namedJdg{\mstrSemExpPred{\Hst}{bs}{e::ck}{\hist{x}}}
        {\mstrEqnPred{G}{\Hst}{bs}{\eqn{x}{e}{ck}}}{NSeq} 
\\ 
   \namedJdg{\predSet{\mstrSemExpPred{\Hst}{bs}{e::ck}{[vs]}}
            {\semFby{c}{vs}{\hist{x}}}}
        {\mstrEqnPred{G}{\Hst}{bs}{\eqn{x}{\fby{c}{e}}{ck}}}{NSfby'} \\
\\     
    \namedJdg{\predSet{\mstrSemExpPred{\Hst}{bs}{\vv{e}}{\tupstrm{vs}}}
    {\mstrSemCkPred{\Hst}{bs}{ck}{\baseOf{\tupstrm{vs}}}}
    {\mstrCallPred{G}{\liftnode{f}}{\tupstrm{vs}}{\vv{\hist{x_i}}}}}
            {\mstrEqnPred{G}{\Hst}{bs}
        {\eqn{\vv{x}}{f(\vv{e})}{ck}  }}{NSncall'}
   
   \end{array}%
    $$
    
    \caption{Stream semantics of \nlustre\ nodes and equations}
    \label{FIG:APPStrSemNodeEqn}
\end{figure*}

The (NSndef') rule only differs from (LSndef) in that 
post-transcription clock alignment, we have an additional requirement of $\Hst$ being in accordance with the base clock $bs$, enforced by \ckFont{respects-clock}.
The (NSeq) rule for simple equations mentions the clock that annotates the defining expression, checking that it is consistent with the assumed history for the defined variable $x$.
The (NSfby') rule for \ttt{fby} in an equational context uses the semantic operation \ckFont{fby}, which differs from $\ckFont{fby}_L$ in that it requires its first argument to be a constant rather than a stream.
Finally, the rule rule (NSncall') for node call, now in an equational context, is similar to (LSncall) except that it constrains the clock modulating the equation to be the base clock of the input flows.

%% file: APP_aux_pred.tex
\section{Auxiliary Predicate Definitions}
\label{APP:AuxPred}
The definitions of the auxiliary semantic stream predicates \ckFont{when}, \ckFont{const}, \ckFont{merge}, \ckFont{ite} are given in \autoref{FIG:Def1}. All predicates except $\textcolor{blue}{\textsf{fby}_L}$ and $\textcolor{blue}{\textsf{fby}_{NL}}$ (defined in \autoref{FIG:LusFbyPred}) are reused to define semantics of both \lustre\ and \nlustre.

\begin{figure*}[ht]
\[
    \begin{array}{c}
        \namedJdg{\semConst{bs'}{c}{cs'}}
        {\semConst{\cc{\true}{bs'}}{c}{\ccnb{\stream{c}}{cs'}}}{DcnstT} ~~~~
        \namedJdg{\semConst{bs'}{c}{cs'}}
            {\semConst{\cc{\false}{bs'}}{c}{\ccnb{\nullStream}{cs'}}}{DcnstF} 
        \\ \\
        \namedJdg{\predSet{\liftunop{es'}{os'}}{\val{v'}= \unop{v}}}
            {\liftunop{\cc{\stream{v}}{es'}}{
                \ccnb{\stream{v'}}{os'}}}{Dunop}  
        ~~~~
        \namedJdg{\liftunop{es'}{os'}}
                {\liftunop{\cc{\nullStream}{es'}}{
                    \ccnb{\nullStream}{os'}}}{DunopA} 
                \\ \\
\namedJdg{es = \ccnb{v}{es'}}
    {\tl{es} = es'}{Dtl}    
~~~~
        \namedJdg{
            \predSet{\liftbinop{es_1'}{es_2'}{os'}}{\binop{v_1}{v_2}=\val{v}}}
        {\liftbinop{\cc{\stream{v_1}}{es_1'}}{\cc{\stream{v_2}}{es_2'}}{
            \ccnb{\stream{v}}{os'}}}{Dbinop}
            \\ \\
    \namedJdg{x \in \tti{dom}(\Hst)}
    {\htl{\Hst}(x) =  \tl{\hist{x}}}{Dhtl}
    ~~~~    

        \namedJdg{\liftbinop{es_1'}{es_2'}{os'}}
        {\liftbinop{\cc{\nullStream}{es_1'}}{\cc{\nullStream}{es_2'}}{
            \ccnb{\nullStream}{os'}}}{DbinopA} 
   \\ \\
   \namedJdg{\resClk{\Hst}{bs}{vs}}
    {\resClk{\Hst}{\cc{\false}{bs}}{\cc{\nullStream}{vs}}}{DresA} 
    ~~~
     \namedJdg{\baseOf{vs}= ~bs}
    {\baseOf{\cc{\val{v}}{vs}} = ~\ccnb{\true}{bs} }{Dbase1}  
     \\ \\
    \namedJdg{\resClk{\Hst}{bs}{vs}}
    {\resClk{\Hst}{\cc{\true}{bs}}{\cc{\stream{v}}{vs}}}{Dres} 
    ~~~~
    \namedJdg{\baseOf{vs}= ~bs}
    {\baseOf{\cc{\nullStream}{vs}} = ~\ccnb{\false}{bs} }{Dbase2}     
    \end{array}
\]
    \caption{Definitions of auxiliary predicates-1}
    \label{FIG:Def1}
\end{figure*}

\begin{figure*}
    \[
    \begin{array}{c}
        \namedJdg{\semMerge{xs'}{ts'}{fs'}{os'}}
        {\semMerge{
            \cc{\stream{T}}{xs'}}
            {\cc{\stream{v_t}}{ts'}}
            {\cc{\nullStream}{fs'}}
            {\ccnb{\stream{v_t}}{os'}}}{DmrgT} 
            \\   \\
        \namedJdg{\semMerge{xs'}{ts'}{fs'}{os'}}
        {\semMerge{
            \cc{\stream{F}}{xs'}}
            {\cc{\nullStream}{ts'}}
            {\cc{\stream{v_f}}{fs'}}
            {\ccnb{\stream{v_f}}{os'}}}{DmrgF} 
            \\ \\
        \namedJdg{\semMerge{xs'}{ts'}{fs'}{os'}}
        {\semMerge{
            \cc{\nullStream}{xs'}}
            {\cc{\nullStream}{ts'}}
            {\cc{\nullStream}{fs'}}
            {\ccnb{\nullStream}{os'}}}{DmrgA} 
            \\ \\
        \namedJdg{\semIte{es'}{ts'}{fs'}{os'}}
        {\semIte{
            \cc{\stream{T}}{es'}}
            {\cc{\stream{v_t}}{ts'}}
            {\cc{\stream{v_f}}{fs'}}
            {\ccnb{\stream{v_t}}{os'}}}{DiteT}  
            \\ \\
            \namedJdg{\semIte{es'}{ts'}{fs'}{os'}}
            {\semIte{
                \cc{ \stream{F}}{es'}}
                {\cc{\stream{v_t}}{ts'}}
                {\cc{\stream{v_f}}{fs'}}
                {\ccnb{\stream{v_f}}{os'}}}{DiteF} 
                \\ \\
            \namedJdg{\semIte{es'}{ts'}{fs'}{os'}}
                 {\semIte{  \cc{\nullStream}{es'}}
                    {\cc{\nullStream}{ts'}}
                    {\cc{\nullStream}{fs'}}
            {\ccnb{\nullStream}{os'}}}{DiteA} \\ 
            \\
                            \namedJdg{\whenk{k}{xs'}{es'}{os'}}
            {\whenk{k}{\cc{\stream{\neg k}}{xs'}}{\cc{\stream{v}}{es'}}{\ccnb{\nullStream}{os'}}}{DwhA1} 
            \\ \\
                \namedJdg{\semFby{v}{xs}{ys}}
            {\semFby{c}{\cc{\stream{v}}{xs}}{\ccnb{\stream{c}}{ys}}}{Dfby} 
            ~~
            \namedJdg{\whenk{k}{xs'}{es'}{os'}}
            {\whenk{k}{\cc{\stream{k}}{xs'}}{\cc{\stream{v}}{es'}}{\ccnb{\stream{v}}{os'}}}{Dwhk}
            \\ \\
            \namedJdg{\semFby{c}{xs}{ys}}
            {\semFby{c}{\cc{\nullStream}{xs}}{\ccnb{\nullStream}{ys}}}{DfbyA} 
            ~~
            \namedJdg{\whenk{k}{xs'}{es'}{os'}}
            {\whenk{k}{\cc{\nullStream}{xs'}}{\cc{\nullStream}{es'}}{\ccnb{\nullStream}{os'}}}{DwhA2} 
         \\ \\
    \end{array}
    \]
    \caption{Definitions of auxiliary predicates-2}
\end{figure*}

All auxiliary stream operators are defined to behave according to the clocking regime.
For example, the rule (DcnstF) ensures the absence of a value when the clock is \ckFont{false}.
Likewise the unary and binary operators lifted to stream operations \ckFont{$\hat{\diamond}$} and \ckFont{$\hat{\oplus}$} operate only when the argument streams have values present, as in (Dunop) and (Dbinop), and mark absence when the argument streams' values are absent, as shown in (DunopA) and (DbinopA).
The rules (Dtl) and (Dhtl) are obvious.

Note that in the rules (DmrgT) and (DmrgF) for \ckFont{merge}, a value is present on one of the two streams being merged and absent on the other.
When a value is absent on the stream corresponding to the boolean variable, values are absent on all streams (DmrgA).
The rules for \ckFont{ite} require all streams to have values present, \textit{i.e.},  (DiteT) and (DiteF), or all absent, \textit{i.e.}, (DiteA).
We have already discussed the \ckFont{when} operation in some detail earlier.

The $\textcolor{blue}{\textsf{fby}_{NL}}$ operation is a bit subtle, and rule (Dfby) may look non-intuitive.  
However, its formulation corresponds exactly to the V\'{e}lus formalisation,  ensuring that a value from the first argument stream is prepended exactly when a leading value would have been present on the second argument stream. 
The operation \ckFont{base-of} converts a value stream to a clock, \textit{i.e.}, a boolean stream.
The operation \ckFont{respects-clock} is formulated corresponding to the V\'{e}lus definition. \\

The main difference between $\textcolor{blue}{\textsf{fby}_L}$ and $\textcolor{blue}{\textsf{fby}_{NL}}$ is that the former takes a \tbf{stream} while the latter takes a constant value. 
The $\textcolor{blue}{\textsf{fby}_{NL}}$ predicate assigns the currently saved constant as the stream value and delays the current operand stream by storing its current value for the next clock cycle (effectively functioning as an initialized D-flip flop).
$\textcolor{blue}{\textsf{fby}_L}$ on the other hand extracts the $0^{th}$ tick value of the first operand stream and  uses the  predicate $\textcolor{blue}{\textsf{fby}_{dl}}$ for delaying.

\begin{figure*}
    $$
    \begin{array}{c}
      \mjdg{\semLFby{xs}{ys}{os}}
      {\semLFby{(\ccnb{\nullStream}{xs})}{(\ccnb{\nullStream}{ys})}{\ccnb{\nullStream}{os}}}  ~~~~
      \mjdg{\semLDFby{y}{xs}{ys}{os}}
      {\semLFby{(\ccnb{\stream{x}}{xs})}{(\ccnb{\stream{y}}{ys})}{ \ccnb{\stream{x}}{os}}}\\
      \\
      \mjdg{\semLDFby{v}{xs}{ys}{os}}
      {\semLDFby{v}{(\ccnb{\nullStream}{xs})}{(\ccnb{\nullStream}{ys})}{\ccnb{\nullStream}{os}}} ~~~~
      \mjdg{\semLDFby{y}{xs}{ys}{os}}
      {\semLDFby{v}{(\ccnb{\stream{x}}{xs})}{(\ccnb{\stream{y}}{ys})} {\ccnb{\stream{v}}{os}}}
    \end{array}
    $$
    \caption{\lustre's \ckFont{fby} semantic predicates}
    \label{FIG:LusFbyPred}
\end{figure*}

%% file: running_example.tex
\section{Example: A More Detailed Analysis}
\label{APP:AnalysisExample}

We adapt the examples given in \cite{Bourke-jfla2021} of translation from \lustre\ to \nlustre, and show how our typing rules and security analysis works.  
We also illustrate the preservation of the security types during the translation.
We annotate the programs with \textcolor{blue}{security types} (as superscripts)
and \textcolor{blue}{constraints} on them for each equation (as comments), according to the typing rules. 

\begin{figure*}
\footnotesize
    \begin{minipage}{.5\textwidth}
\begin{lstlisting}[language=Lustre]
node cnt_dn
  (res$^{\type{\alpha_1}}$: bool; n$^{\type{\alpha_2}}$: int) 
  returns (cpt$^{\type{\beta}}$: int); 
let
  (cpt$^{ck}$)$^{\type{\beta^\gamma}}$ = if res$^{\type{\alpha_1}}$ then n$^{\type{\alpha_2}}$ 
    else (n$^{\type{\alpha_2}}$ fby (cpt$^{\type{\beta}}$-1));    
  -- $\type{\rho_L} =  \{ \type{ \gamma \lub \alpha_1 \lub \alpha_2 \lub \alpha_2 \lub \beta \lub \bot \strel \beta} \} $ 
tel
\end{lstlisting}
\hrule
\begin{flushleft}
  \footnotesize
    \begin{align*}
{}&\lsimpl{(\type{\beta}, \type{ \rho_L } )}{\{\}} =  
~(\type{\beta}, \{ \type{\gamma \lub \alpha_1 \lub \alpha_2 ~\strel~ \beta } \}) \\
{}&\nlsimpl{ (\type{\beta}, (\type{\rho_1} \cup \type{\rho_2} \cup \type{\rho_3} \cup \type{\rho_4}) )}{\vv{\delta}} \\
{}& = ~~ (\type{\beta}, \{ \type{\gamma \lub \alpha_1 \lub \alpha_2 ~\strel~ \beta } \}) \\
{}&\textit{where} ~ \type{\vv{\delta}} =  ~~\{\{\type{\delta_1},\type{\delta_2},\type{\delta_3}\}\} 
\end{align*}
\end{flushleft}
    \end{minipage}%
    \begin{minipage}{.5\textwidth}
\begin{lstlisting}[language=Lustre]
node cnt_dn
  (res$^{\type{\alpha_1}}$: bool; n$^{\type{\alpha_2}}$: int) 
  returns (cpt$^{\type{\beta}}$: int); 
  var v14$^{\type{\delta_1}}$, v24$^{\type{\delta_2}}$, v25$^{\type{\delta_3}}$:int; 
let
  v24$^{\type{\delta_2}}$ =$^{\type{\gamma}}$ true fby false;
  -- $\type{\rho_1} = \{ \type{\gamma \lub \bot \lub \bot \rel \delta_2} \}$
  v25$^{\type{\delta_3}}$ =$^{\type{\gamma}}$ 0 fby (cpt$^{\type{\beta}}$ -1);
  -- $\type{\rho_2} = \{ \type{\gamma \lub \beta \lub \bot \rel \delta_3} \}$
  v14$^{\type{\delta_1}}$ =$^{\type{\gamma}}$ if v24$^{\type{\delta_2}}$ then n 
    else v25$^{\type{\delta_3}}$;
  -- $\type{\rho_3}  = \{ \type{\gamma \lub \delta_2 \lub \delta_3 \rel \delta_1} \}$
  cpt$^{\type{\beta}}$ =$^{\type{\gamma}}$ if res$^{\type{\alpha_1}}$ then n$^{\type{\alpha_2}}$ 
    else v14$^{\type{\delta_1}}$;    
  -- $ \type{\rho_4}  = \{ \type{ \gamma \lub \alpha_1 \lub \alpha_2 \lub \delta_1 \strel \beta} \}$ 
tel
\end{lstlisting}
    \end{minipage}

    \noindent\rule{\linewidth}{0.4pt}
    
    \caption{Example of normalisation with security analysis}
    \label{FIG:L2NLEx1}
    \end{figure*}
    
The node \ttt{cnt\_dn} implements a count down timer \ttt{cpt} which is initialized with the value of \ttt{n} on $0^{th}$ tick and whenever there is a \ttt{T} on reset \ttt{res}. 
Changing the value of \ttt{n} when the reset is \ttt{F} doesn't affect the count.

We assign security types $\type{\alpha_1}$ to
input \ttt{res}, and $\type{\alpha_2}$ to input \ttt{n}.
The output \ttt{cpt} is assigned security type $\type{\beta}$, and the clock $ck$ the type $\type{\gamma}$.
There are no local variables.
Based on the rules (LTvar), (LTbinop), (LTfby) and (LTite), we get constraint $\type{\rho_L}$.
After simplification, the resultant security signature of \ttt{cnt\_dn} is given by:
\[
  \securitySignature{\ttt{cnt\_dn}}
  {(\type{\alpha_1,\alpha_2})}
  {\type{\beta}}
  {\type{\gamma}}
  {\{ \type{\gamma \stlub \alpha_1 \stlub \alpha_2 ~\strel~ \beta} \}}
\]

The normalisation pass de-nests the \ttt{fby} expression and explicitly initializes it into 3 different local streams (\ttt{v14},\ttt{v24},\ttt{v25}).
These have security types
$\type{\delta_1}, \type{\delta_2}, \type{\delta_3}$.
The local variables generate constraints $\type{\rho_1}, \type{\rho_2}, \type{\rho_3}$ which are eliminated by \ckFont{simplify}.

The resultant signature of \ttt{cnt\_dn} in the translated program is also given by:
\[
  \securitySignature{\ttt{cnt\_dn}}
  {(\type{\alpha_1,\alpha_2})}
  {\type{\beta}}
  {\type{\gamma}}
  {\{ \type{\gamma \stlub \alpha_1 \stlub \alpha_2 ~\strel~ \beta} \}}
\]


The \ttt{re\_trig} node in  \autoref{FIG:L2NLEx2} uses the \ttt{cnt\_dn} node (\autoref{FIG:L2NLEx1}) to implement a count-down timer that is explicitly triggered whenever there is a rising edge (represented by \ttt{edge})  on \ttt{i}. 
If the count \ttt{v} expires to $0$ before a \ttt{T} on \ttt{i}, the counter isn't allowed restart the count. 
Output \ttt{o} represents an active count in progress. 

\ignore{
We annotate the program with \textcolor{blue}{security types} (superscripts)
and \textcolor{blue}{constraints} on them for each equation (as comments), according to the typing rules.
\ttt{cnt\_dn} is assumed to have security signature
\[
  \securitySignature{\ttt{cnt\_dn}}
  {(\type{\alpha_1,\alpha_2})}
  {\type{\beta}}
  {\type{\gamma}}
  {\{ \type{\gamma \stlub \alpha_1 \stlub \alpha_2 ~\strel~ \beta} \}}
\]
}
Eliminating the security types $\type{\delta'_1}, \type{\delta'_2}, \type{\delta'_3}$,  and $\type{\delta'_6}$, 
of the local variables \ttt{edge}, \ttt{c}, \ttt{v} and nested call to \ttt{cnt\_dn} respectively, we
get the constraint $\{ \type{\gamma' \lub \alpha'_1 \lub \alpha'_2 \strel \beta'}\}$.

Normalisation introduces local variables (\ttt{v21,v22,v24}) with security types
$\type{\delta'_4}, \type{\delta'_5},  \type{\delta'_6}$.
(Identical names have been used to show the correspondence.)
The $\type{\delta'_i}$ are eliminated by \ckFont{simplify}, and the refinement type $\type{\delta'_6 \{\!| \rho' |\!\}}$ for the node call in the \lustre\ version becomes an explicit constraint $\type{\rho_5}$ in \nlustre.
We see that the security signature of \ttt{re\_trig} remains the same.

\begin{figure*}
\footnotesize
    \begin{minipage}{.5\textwidth}
        \begin{lstlisting}[language=Lustre]
node re_trig(i$^{\type{\alpha'_1}}$:bool; n$^{\type{\alpha'_2}}$:int)
  returns (o$^{\type{\beta'}}$ : bool)
  var edge$^{\type{\delta'_1}}$, c$^{\type{\delta'_2}}$:bool,
   v$^{\type{\delta'_3}}$:int;
let
  (edge$^{ck}$)$^{\type{\delta^{'\gamma'}_1}}$ = i$^{\type{\alpha'_1}}$ and 
    (false$^{\type{\bot}}$ fby (not i$^{\type{\alpha'_1}}$)); 
-- $\type{\rho_{1L}}= \{\type{\bot \lub \alpha'_1 \lub \bot \lub \alpha'_1 \rel \delta'_1}\}$
  (c$^{ck}$)$^{\type{\delta^{'\gamma'}_2}}$ = edge$^{\type{\delta'_1}}$ or 
    (false$^{\type{\bot}}$ fby o$^{\type{\beta'}}$);
-- $\type{\rho_{2L}}= \{\type{\gamma' \lub \delta'_1 \lub \bot \lub \beta' \rel \delta'_2}\}$
  (v$^{c}$)$^{\type{\delta^{'\delta'_2}_3}}$ = merge c$^{\type{\delta'_2}}$
   (cnt_dn((edge$^{\type{\delta'_1}}$, n$^{\type{\alpha'_2}}$)
    when c$^{\type{\delta'_2}}$))$^{{\type{\delta'_6\{|\rho'|\}}}^{\type{\delta'_2}}}$ 
   (0 when not c$^{\type{\delta'_2}}$);
-- $\type{\rho'}= \{ \type{\delta'_2 \lub (\delta'_1 \lub \delta'_2) \lub (\alpha'_2 \lub \delta'_2) \rel\delta'_6}\}$
-- $\type{\rho_{3L}}= \{\type{\delta'_2 \lub \delta'_2 \lub \delta'_6 \lub \bot \lub \delta'_2 \rel \delta'_3}\} \cup \type{\rho'}$
   (o$^{c}$)$^{\type{\beta^{'\delta'_2}}}$= v$^{\type{\delta'_3}}$ > 0$^{\type{\bot}}$;
-- $\type{\rho_{4L}}= \{\type{\delta'_2 \lub \delta'_3 \lub \bot \rel \beta'}\}$
tel
    \end{lstlisting}
    \end{minipage}%
        \begin{minipage}{.5\textwidth}

\begin{lstlisting}[language=Lustre]
node re_trig(i$^{\type{\alpha'_1}}$:bool; n$^{\type{\alpha'_2}}$:int)
 returns (o$^{\type{\beta'}}$ : bool)
 var edge$^{\type{\delta'_1}}$, ck$^{\type{\delta'_2}}$:bool, v$^{\type{\delta'_3}}$:int,
  v22$^{\type{\delta'_4}}$:bool, v21$^{\type{\delta'_5}}$:bool,
  v24$^{\type{\delta'_6}}$:int when ck;
let
  v22$^{{\type{\delta_4}}}$  =$_{\type{\delta'_2}}$ false$^{\type{\bot}}$ fby 
    (not i$^{\type{\alpha'_1}}$);
-- $\type{\rho_1} = \{\type{\delta'_2 \lub \bot \lub \alpha'_1 \rel \delta'_4}\} $ 
  edge$^{\type{\delta'_1}}$ =$_{\type{\bot}}$ i$^{\type{\alpha'_1}}$ and v22$^{\type{\delta'_4}}$;
-- $\type{\rho_2} = \{\type{ \bot \lub \alpha'_1 \lub \delta'_4 \rel \delta'_1}\} $ 
  v21$^{\type{\delta'_5}}$ =$_{\type{\bot}}$ false$^{\type{\bot}}$ fby o$^{\type{\beta'}}$;
-- $\type{\rho_3} = \{\type{\bot \lub \bot \lub \beta' \rel \delta'_5}\}$
  ck$^{\type{\delta'_2}}$  =$_{\type{\gamma'}}$ edge$^{\type{\delta'_1}}$ or v21$^{\type{\delta'_5}}$;
-- $\type{\rho_4} = \{\type{\bot \lub \delta'_1 \lub \delta'_5 \rel \delta'_2}\}$
  v24$^{\type{\delta'_6}}$ =$_{\type{\delta'_2}}$ cnt_dn(
    edge$^{\type{\delta'_1}}$ when ck$^{\type{\delta'_2}}$,
   n$^{\type{\alpha'_2}}$ when ck$^{\type{\delta'_2}}$);
-- $\type{\rho_5} = \{\type{\delta'_2 \lub (\delta'_1 \lub \delta'_2) \lub (\alpha'_2 \lub \delta'_2) \rel \delta'_6}\}$
  v$^{\type{\delta'_3}}$ =$_{\type{\delta'_2}}$ merge ck$^{\type{\delta'_2}}$ v24$^{\type{\delta'_6}}$ 
    (0$^{\type{\bot}}$ when not ck$^{\type{\delta'_2}}$);
-- $\type{\rho_6}= \{\type{\delta'_2 \lub \delta'_2 \lub \delta'_6 \lub \bot \lub \delta'_2 \rel \delta'_3}\}$
  o$^{\type{\beta'}}$ =$^{\type{\delta'_2}}$ v$^{\type{\delta'_3}}$>0$^{\type{\bot}}$;
-- $\type{\rho_7}= \{\type{\delta'_2 \lub \delta'_3 \lub \bot \rel \beta'}\}$
tel
\end{lstlisting}
        \end{minipage}
        \noindent\rule{\linewidth}{0.4pt}
        \centering
        \begin{align*}
        \lsimpl{(\type{\beta'}, \{ \union{\type{\rho_{1L}}}{\type{\rho_{2L}}}{\type{\rho_{3L}}}{\type{\rho_{4L}}}\})}{\{\type{\delta'_1},\type{\delta'_2},\type{\delta'_3}, \type{\delta'_6}\}} = (\type{\beta'}, \{ \type{\gamma' \lub \alpha'_1 \lub \alpha'_2 \rel \beta'}\}) \\
        \nlsimpl{(\type{\beta'}, \{ \union{\type{\rho_{1}}}{\type{\rho_{2}}}{\type{\rho_{3}}}{\type{\rho_{4}}}{\type{\rho_{5}}}{\type{\rho_{6}}}{\type{\rho_{7}}}\})}{\{\type{\delta'_1},\type{\delta'_2},\type{\delta'_3}, \type{\delta'_4}, \type{\delta'_5}, \type{\delta'_6}\}} \\
        \hfill = (\type{\beta'},\{ \type{\gamma' \lub \alpha'_1 \lub \alpha'_2 \rel \beta'}\})
      \end{align*}
        \caption{Example: Security analysis and normalisation}
        \label{FIG:L2NLEx2}
\end{figure*}